\documentclass[12pt]{article}
\pdfoutput=1 
\usepackage{tikz}
\usepackage{amssymb}
\usepackage{epsfig}
\usepackage{bbm}
\usepackage{bbold}
\usepackage{graphicx}
\usepackage{hyperref}
\usepackage{romannum}
\usepackage{graphics,graphpap}
\usepackage{amsmath,exscale,amsthm}
\usepackage{graphicx}
\usepackage{yfonts}
\usepackage{mathrsfs}
\usetikzlibrary{trees}
\usetikzlibrary{positioning,arrows}
\usetikzlibrary{decorations.pathmorphing}
\usetikzlibrary{decorations.markings}
\usetikzlibrary{decorations.text}
\usetikzlibrary{fit,chains,calc,shapes.geometric,intersections}
\usepackage[none]{hyphenat}
\usetikzlibrary{matrix,positioning,decorations.pathreplacing}

\usepackage{amsfonts}

\def\L{\mathcal L}
\def\e{\varepsilon}

\newcommand{\wt}{\widetilde}

\begin{document}

\def\a{\alpha}
\def\b{\beta}
\def\c{\chi}
\def\d{\delta}
\def\e{\epsilon}
\def\f{\phi}
\def\g{\gamma}
\def\h{\eta}
\def\i{\iota}
\def\j{\psi}
\def\k{\kappa}
\def\la{\lambda}
\def\m{\mu}
\def\n{\nu}
\def\o{\omega}
\def\p{\pi}
\def\q{\theta}
\def\r{\rho}
\def\s{\sigma}
\def\t{\tau}
\def\u{\upsilon}
\def\x{\xi}
\def\z{\zeta}
\def\D{\Delta}
\def\F{\Phi}
\def\G{\Gamma}
\def\J{\Psi}
\def\L{\Lambda}
\def\O{\Omega}
\def\P{\Pi}
\def\Q{\Theta}
\def\S{\Sigma}
\def\U{\Upsilon}
\def\X{\Xi}

\def\ve{\varepsilon}
\def\vf{\varphi}
\def\vr{\varrho}
\def\vs{\varsigma}
\def\vq{\vartheta}

\def\dg{\dagger}                                     
\def\ddg{\ddagger}                                   
\def\wt#1{\widetilde{#1}}                    
\def\mt{\widetilde{m}_1}
\def\mti{\widetilde{m}_i}
\def\mtj{\widetilde{m}_j}
\def\rt{\widetilde{r}_1}
\def\mtt{\widetilde{m}_2}
\def\mttt{\widetilde{m}_3}
\def\rtt{\widetilde{r}_2}
\def\mb{\overline{m}}
\def\VEV#1{\left\langle #1\right\rangle}        
\def\be{\begin{equation}}
\def\ee{\end{equation}}
\def\ds{\displaystyle}
\def\ra{\rightarrow}

\def\bea{\begin{eqnarray}}
\def\eea{\end{eqnarray}}
\def\NO{\nonumber}
\def\Bar#1{\overline{#1}}


\def\pl#1#2#3{Phys.~Lett.~{\bf B {#1}} ({#2}) #3}
\def\np#1#2#3{Nucl.~Phys.~{\bf B {#1}} ({#2}) #3}
\def\prl#1#2#3{Phys.~Rev.~Lett.~{\bf #1} ({#2}) #3}
\def\pr#1#2#3{Phys.~Rev.~{\bf D {#1}} ({#2}) #3}
\def\zp#1#2#3{Z.~Phys.~{\bf C {#1}} ({#2}) #3}
\def\cqg#1#2#3{Class.~and Quantum Grav.~{\bf {#1}} ({#2}) #3}
\def\cmp#1#2#3{Commun.~Math.~Phys.~{\bf {#1}} ({#2}) #3}
\def\jmp#1#2#3{J.~Math.~Phys.~{\bf {#1}} ({#2}) #3}
\def\ap#1#2#3{Ann.~of Phys.~{\bf {#1}} ({#2}) #3}
\def\prep#1#2#3{Phys.~Rep.~{\bf {#1}C} ({#2}) #3}
\def\ptp#1#2#3{Progr.~Theor.~Phys.~{\bf {#1}} ({#2}) #3}
\def\ijmp#1#2#3{Int.~J.~Mod.~Phys.~{\bf A {#1}} ({#2}) #3}
\def\mpl#1#2#3{Mod.~Phys.~Lett.~{\bf A {#1}} ({#2}) #3}
\def\nc#1#2#3{Nuovo Cim.~{\bf {#1}} ({#2}) #3}
\def\ibid#1#2#3{{\it ibid.}~{\bf {#1}} ({#2}) #3}

\title{{\bf 
Density matrix calculation of the dark matter abundance in the Higgs induced right-handed neutrino mixing model
}
\author{\large P. Di Bari$^{a}$,  K. Farrag$^{a,b}$, 
R. Samanta$^{a}$ and Y.L. Zhou$^{a}$ \\
$^a${\it\small School of Physics and Astronomy},
{\it\small University of Southampton,} 
{\it\small  Southampton, SO17 1BJ, U.K.} \\
$^b${\it\small School of Physics and Astronomy}, 
{\it\small Queen Mary, University of London} 
{\it\small  London, E1 4NS, U.K.}
}}

\maketitle \thispagestyle{empty}

\vspace{-11mm}

\pagenumbering{arabic}
\begin{abstract}
We present new results on the calculation of the dark matter relic abundance
within the Higgs induced right-handed (RH) neutrino mixing model, solving the  associated density matrix equation.
For a benchmark value of the dark matter mass $M_{\rm DM} = 220\,{\rm TeV}$, 
we show the evolution of the abundance and how this depends  on  reheat temperature, 
dark matter lifetime 
and source RH neutrino mass $M_{\rm S}$, with the assumption $M_{\rm S} < M_{\rm DM}$.
We compare the results with those obtained within the Landau-Zener approximation, showing that
the latter largely overestimates the final abundance giving some analytical insight. 
However, we also notice that since in the density matrix formalism the production is 
non-resonant, this allows source RH neutrino masses below the W boson mass, making
dark matter more stable at large mass values. This opens an allowed region for initial
vanishing source RH neutrino abundance. 
For example, for $M_{\rm S} \gtrsim 1\,{\rm GeV}$, we find $M_{\rm DM}\gtrsim 20\,{\rm PeV}$.
Otherwise, for $M_{\rm S} > M_W\sim 100\,{\rm GeV}$,
one has to  assume a thermalisation of the source RH neutrinos prior to the freeze-in of the dark matter abundance.
This results into a large allowed range for $M_{\rm DM}$, depending on $M_{\rm S}$. 
For example, imposing $M_{\rm S} \gtrsim 300\,{\rm GeV}$, allowing also
successful leptogenesis, we find $0.5 \lesssim M_{\rm DM}/{\rm TeV} \lesssim 500$.
We also discuss in detail leptogenesis with two quasi-degenerate RH neutrinos,
showing a case when observed dark matter abundance and matter-antimatter asymmetry are 
simultaneously reproduced.
Finally, we comment on how an initial thermal source RH neutrino abundance  can be justified
and on how our results suggest that also the interesting case where $M_{\rm DM} < M_{\rm S}$, 
embeddable in usual high scale two RH neutrino seesaw models, might be viable. 
\end{abstract}

\section{Introduction}

There are different proposals for extending the Standard Model 
in a way to explain neutrino masses and mixing and 
at the same time addressing two of the most compelling cosmological puzzles: 
dark matter (DM) and matter-antimatter asymmetry of the universe. 
Such extensions are traditionally based on new physics at energy scales
inaccessible with ground laboratories and, therefore, usually untestable.  
Moreover, one of course would like to consider models that are as minimal as possible. 
An attractive extension that fulfils both conditions, testability and minimality,
and that provides a unified picture of neutrino masses and mixing, dark matter and the matter-antimatter asymmetry 
of the universe with leptogenesis is the scenario of (cold) dark matter from
Higgs induced right-handed neutrino mixing (Higgs induced RHiNo DM) \cite{ad,unified}.
This is based, in addition to a traditional type-I seesaw Lagrangian extension of the SM with right-handed (RH) neutrinos,  
on the introduction of a non-renormalizable 5-dim operator, 
\be\label{anisimov}
{\cal O}_A = {\la_{IJ} \over \L} \, \Phi^\dagger \, \Phi \, \overline{N_I^c} \, N_J \,  ,
\ee
coupling the standard Higgs doublet to RH neutrinos and inducing a RH-RH neutrino mixing, therefore, differently
from the usual RH-LH neutrino mixing already rising from the type-I seesaw Lagrangian. We will refer to this operator 
as the Anisimov operator \cite{anisimov,bezrukov,ad}. It can be regarded as a special case
of Higgs portal operator, though not strictly falling within the categories considered in \cite{wilczek}.

The interesting feature of the Anisimov operator is that in addition to allow the production 
of a decoupled RH neutrino playing the role of DM particle, it also predicts a contribution from RH neutrino DM decays
to the  flux of very high energy neutrinos detectable at neutrino telescopes \cite{ad}.
Therefore, the recent IceCube neutrino telescope discovery of a very high energy neutrino component  
in excess of the  well known atmospheric contribution \cite{icecube1,icecube2,icecube3,icecube4,icecube5}, 
prompts the question whether, in addition to an expected, though yet
largely undetermined, astrophysical component, the IceCube signal might also receive a contribution of cosmological origin from DM decays. Initial analyses mainly focused on a  scenario where heavy DM decays can explain
the whole signal and in particular an excess of PeV neutrinos in early data \cite{feldskusyana,serpico,murase}.  
This possibility seems now  disfavoured by the latest data \cite{icecube6}, though not completely excluded \cite{sergio}.
However, current IceCube data on the energy spectrum favour the presence of an 
extra contribution in addition to a 
traditional astrophysical component described by a power law 
with spectral index $\gamma \simeq 2$, as predicted by the Fermi mechanism \cite{dev}.  
Different analyses have shown that in particular the addition of 
a contribution from DM decays can help in explaining the IceCube 
data \cite{chianese1,chianese2,chianese3,dev2,sergio}. In particular, Higgs induced 
RHiNo DM also provides a good fit to the data
for  DM masses  in the range $\sim 100\,{\rm TeV}$--$1\,{\rm PeV}$ \cite{unified}.  
After the IceCube discovery of very high energy neutrinos, various models 
have been presented that could potentially produce an excess with respect to an astrophysical component \cite{manymodels1,manymodels2,manymodels3}. 
However, the Higgs induced RHiNo DM 
has the attractive features of minimality and also predictivity,  since
the same interactions, described by the Anisimov operator (\ref{anisimov}), 
can be responsible both for DM  production and for its decays.  
Evidence of this predictive power is that already in the original proposal, prior to IceCube discovery,
it was pointed out how the allowed range of DM masses 
could be probed by neutrino telescopes and in particular by IceCube \cite{ad}.
At the same time the model explains neutrino masses and mixing within
a traditional type-I seesaw mechanism and the matter-antimatter asymmetry of the universe with 
leptogenesis. In this way a unified picture of neutrino masses, dark matter and leptogenesis, satisfying all experimental constraints, 
is possible within a certain region in the DM mass-lifetime plane \cite{unified}.

This intriguing phenomenological picture  provides a strong motivation for a more solid calculation of the 
Higgs induced  RHiNo DM  relic abundance, a key ingredient for the determination of the  allowed mass range.
In \cite{ad,unified} different approximations and simplified assumptions were adopted.
 In particular, a simplistic Landau-Zener (LZ) approximation was used to calculate
the fraction of source RH neutrino abundance that gets non-adiabatically converted into a
DM RH neutrino abundance. In this paper we present results on the 
calculation of the relic DM abundance using the density matrix formalism. 
 
 There are analogies with the calculation of a (light) sterile neutrino abundance 
 from active-sterile neutrino mixing \cite{lzacst} that can lead to a warm DM  
 solution for keV sterile neutrinos \cite{dw}.
 However,  the great difference and complication is that in the case of Higgs induced RHiNo DM
the vacuum mixing angle vanishes and its role is replaced by a misalignment between the Yukawa interactions and the   Higgs induced interactions. This depends on temperature, making the evolution of the system more complicated. In the calculation we still employ a monochromatic approximation and we leave a full momentum dependent calculation for a future investigation. In the 
final discussion we briefly comment on an  extension of the results taking into account momentum dependence.

We also consider the dependence of the relic DM abundance  on the initial conditions, in particular
the dependence on the initial source RH neutrino abundance and on the reheat temperature. 

The paper is structured as follows. In Section 2 we review the model but also improve different 
results such as the lifetime for two body decays
and generalise others. For example, we notice that the four body decays upper bound on $M_{\rm DM}$ does not
apply when the source RH neutrinos are lighter than the $W$ boson.
In Section 3 we introduce the density matrix formalism and the equations we solve.
In Section 4 we show the evolution of the DM abundance and its 
dependence on $T_{\rm RH}, \tau_{\rm DM}$ and  $M_{\rm DM}/M_{\rm S}$
for a benchmark case $M_{\rm DM} =220 \,{\rm TeV}$. We show that the LZ approximation overestimates the
relic abundance by many orders of magnitude and provide some analytic insight that explains this result.
For this benchmark case we also show an example of how observed DM and matter-antimatter asymmetry 
of the universe can be simultaneously reproduced.
 In Section 5 we show the bounds on $M_{\rm DM}$ within different assumptions. 
 In Section 6 we conclude, briefly discussing how the presented results can be extended in different ways.
 Appendix A contains a derivation of the two body decay rate, Appendix B discusses different equivalent ways to write the
 density matrix equation, Appendix C contains an original discussion of leptogenesis with two quasi-degenerate RH neutrinos.
 
\section{Higgs induced RHiNo dark matter and  the LZ  \\ approximation}

Let us briefly review the Higgs induced RHiNo DM model and how the relic DM abundance is calculated 
and DM mass constraints derived within the LZ approximation. 
At the same time we improve and extend some  results on the lifetime of the DM RH neutrino.
The effective Lagrangian  is given by the traditional type-I seesaw Lagrangian \cite{seesaw} with three RH neutrinos 
with the addition of the Anisimov operator.  Before electroweak  spontaneous symmetry breaking one has
($\a=e,\m,\t$ and $I,J=1,2,3$),
\be\label{lagrangian}
-{\cal L}_{M+\L} = \overline{L_{\a}}\,h_{\a J}\, N_{R J}\, \widetilde{\Phi} +
                          \frac{1}{2} \, \overline{N^{c}_{R I}} \, D_{M IJ} \, N_{R J}  +  
                          {\la_{IJ} \over \L} \, \Phi^\dagger \, \Phi \, \overline{N_{R I}^c} \, N_{R J}
                          + \mbox{\rm h.c.}  \,  ,
\ee
where $L_\a^T \equiv (\nu_{L\a},\a_L)$ are the leptonic doublets,
$\Phi$ is the Higgs doublet, with  $\widetilde{\Phi} \equiv {\rm i}\,\sigma_2\,\Phi^{\star}$,  
the $h_{\a J}$'s are the neutrino Yukawa couplings in the {\em flavour basis} where both charged lepton and Majorana mass matrices are diagonal, and 
we defined $D_{M} \equiv {\rm diag}(M_1, M_2, M_3)$, where 
$M_1 \leq M_2 \leq M_3$ are the three heavy neutrino masses. 

After spontaneous symmetry breaking the
Higgs vev generates a neutrino Dirac mass matrix $m_{D} = v \, h$.
One of the three RH neutrinos is assumed to have vanishing Yukawa couplings
so that the entries of one of the three columns in  $h$ and $m_D$ vanish.
This assumption can be justified imposing, for example, a $Z_2$ symmetry. 
For this reason the seesaw formula, 
\be\label{seesaw}
D_m \equiv {\rm diag}(m_1 = 0, m_2, m_3) =  U^{\dagger} \, m_D \, \frac{1}{D_M} \, m_D^T  \, U^{\star} ~,
\ee
where $U$ is the leptonic mixing matrix,  reduces to the two RH neutrino case
with vanishing lightest neutrino mass $m_1 =0$, 
so that the model strictly predicts hierarchical light neutrino masses.
In the Yukawa basis the Yukawa matrix is by definition diagonal and given by 
$D_h\equiv {\rm diag}(h_A, h_B, h_C)$, with $h_A=0 < h_B < h_C$, and the transformation between the two
bases is described by a bi-unitary transformation
\be
m_D = V_L^\dagger \, D_{m_D} \, U_R \,   ,
\ee
where $U_R$ acts on RH neutrino fields and can be regarded as the RH neutrino mixing matrix in the absence
of Higgs induced interactions described by the Anisimov operator for $\la_{IJ} = 0$. 
In this case, the DM RH neutrino Majorana mass eigenstate, that we indicate with $N_{\rm DM}$, 
coincides with the Yukawa eigenstate $N_A$ and is rigorously stable
but also fully decoupled, so that there would be no way to produce it.\footnote{Notice that it does not have
to coincide exactly since it could be long-lived with a life-time $\tau_{DM} \gtrsim 10^{28}\,{\rm s}$
to avoid constraints from IceCube that we discuss later on. In this case, instead of a $m_D$ column with three
texture zeros one would have some very small entries. However, in this case the mixing angles in $U_R$ between
the dark and the other two RH neutrinos would have to 
be so minuscule, and correspondingly the entries in the DM RH neutrino column, 
that for simplicity they can be set to zero for all purposes. Therefore,
the requirement that one mass eigenstate lifetime is sufficiently long-lived to play the role of DM 
evading experimental constraints, implies, within tiny corrections, both vanishing DM RH neutrino Yukawa 
couplings and also vanishing mixing with the other two RH neutrinos, encoded in $U_R$ (this was discussed in detail in \cite{ad}).}
When Higgs induced interactions are switched on, for $\la_{I J} \neq 0$, 
they  trigger a mixing between $N_{\rm DM}$ and the two coupled RH neutrinos.

At finite temperatures, the Yukawa and the Higgs induced interactions 
contribute to the RH neutrino self-energies, producing effective potentials that in general are not 
diagonal in the same basis. The misalignment between the two bases is responsible for  RH neutrino mixing.
For simplicity, but also because this minimises the constraints from decays, as shown in \cite{unified},
it is convenient to assume that the RH neutrino mixing is dominantly
between the DM RH neutrino $N_{\rm DM}$ and only one of the other two RH neutrinos with non-vanishing Yukawa couplings, that we refer to as the source RH neutrino and we indicate  
with $N_{\rm S}$.\footnote{$N_{\rm DM}$ and $N_{\rm S}$ have to be regarded as the
two RH neutrino Majorana mass  eigenstates with $M_{\rm DM} = M_I$ 
and $M_{\rm S} = M_J$ for $I \neq J=1,2,3$. They coincide with the energy eigenstates 
only if $\la_{I J} =0$.}  
In this case we can consider a simple two-neutrino mixing formalism and we indicate the
coupling between the DM and the source RH neutrino with $\lambda_{\rm DM-S}$.
We stress again that for $\lambda_{\rm DM-S} =0$ there would be no mixing, because, as we said, 
$N_{\rm DM}$ would be completely decoupled. Notice also that within this two neutrino mixing formalism, 
Yukawa basis and flavour basis coincide, i.e., $U_R = I$ (see footnote 1).

The Yukawa interactions clearly produce a diagonal contribution to 
the RH neutrino Hamiltonian in the Yukawa basis given by~\cite{weldon}
\be
V^{Y}_{IJ} =  \frac{T^2}{8\,E_J} \, h^2_J \, \d_{IJ} \hspace{3mm} (I,J={\rm DM},{\rm S})  ,
\ee
where $E_J$ is the energy of the Majorana mass eigenstate $N_J$ and
$h_{\rm S} \equiv \sqrt{(h^\dagger \, h)_{\rm SS}}$ is the Yukawa coupling
of $N_{\rm S}$ to the thermal bath (while of course $N_{\rm DM}$ has 
no Yukawa  interactions, since we are assuming $h_{\rm DM} = h_A =0$).
On the other hand, the Higgs induced interactions, described by the Anisimov operator, 
are in general non-diagonal in the Yukawa basis and they produce an effective potential
\be\label{LAMBDAmixing}
V^{\L}_{I J} \simeq \frac{T^2}{12\,\L}\,\la_{I J} \,  \hspace{3mm} (I,J={\rm DM},{\rm S}) .
\ee
%
%
%
In the basis of Majorana mass eigenstates one has also to consider the usual diagonal kinetic 
contribution so that the Hamiltonian  can be written as
\be\label{hamiltonian}
{\cal H}_{IJ} =  \left( \begin{array}{cc}
E_{\rm DM} &  \frac{T^2}{12\,\widetilde{\L}} \\[1ex]
\frac{T^2}{12\,\widetilde{\L}}  &  E_{\rm S} + \frac{T^2}{8\,E_{\rm S}} \, h^2_{\rm S}
\end{array}\right)  \,  ,
\ee
where $\widetilde{\L} \equiv \L/\la_{\rm DM-S}$ and we assumed that the diagonal terms of the Higgs induced interactions  can be neglected. 
Subtracting  a contribution to ${\cal H}$ proportional to the identity, not affecting the mixing, the effective mixing Hamiltonian is then given by
\be\label{effectiveham}
\Delta {\cal H}_{IJ} \simeq   
\left( \begin{array}{cc}
- \frac{\D M^2}{4 \, p} - \frac{T^2}{16\,p} \, h^2_{\rm S} &  \frac{T^2}{12\,\widetilde{\L}}  \\[1ex]
\frac{T^2}{12\,\widetilde{\L}} &  \frac{\D M^2}{4 \, p} + \frac{T^2}{16 \, p} \, h^2_{\rm S}  
\end{array}\right)  \, ,
\ee
where we defined $\D M^2 \equiv M^2_{\rm S} - M^2_{\rm DM}$.  If we adopt a monochromatic approximation, 
so that the momentum is replaced by its average value $p \simeq 3\,T$, 
the effective mixing Hamiltonian in the flavour basis becomes
\be
\Delta {\cal H}_{IJ} \simeq   
\frac{\D M^2}{12\,T}\,\left( \begin{array}{cc}
	- 1 - v_{\rm S}^Y &  \sin 2\theta_{\L}  \\[1ex]
\sin 2\theta_{\L}   &  1 + v_{\rm S}^Y 
\end{array}\right)  \,  .
\ee
Here  we have also introduced the dimensionless effective potential $v_{\rm S}^Y \equiv T^2\,h^2_{\rm S} / (4\,\D M^2)$ and the effective mixing angle $\sin 2\theta_{\L}(T) \equiv T^3/(\widetilde{\L} \, \D M^2 )$, 
that, as we said, is produced by the misalignment between the Yukawa and the Higgs induced interactions.\footnote{In the standard neutrino mixing case, among left handed neutrinos, this effective mixing angle would correspond to  the vacuum mixing angle.}  

If $M_{\rm DM} > M_{\rm S}$, implying $\Delta M^2 <0$, there is a resonance for $v_S^Y = -1$, corresponding
to a specific value of the temperature, the {\em resonance temperature}, given by
\be
T_{\rm res}  \equiv \frac{2 \, \sqrt{|\D M^2|}}{h_{\rm S} } =  \frac{2 \, \sqrt{ M^2_{\rm DM} - M^2_{\rm S}}}{h_{\rm S}} \,  .
\ee
Since the process is highly 
non-adiabatic,\footnote{This means that the neutrino states, that are initially produced as source RH neutrinos
by Yukawa interactions, do not track the matter eigenstates $N_{\Romannum{1}}^{\rm m}(T)$
and $N_{\Romannum{2}}^{\rm m}(T)$ given explicitly by
\bea\label{eneigen}
N_{\Romannum{1}}^{\rm m}(T) & = & N_{\rm DM} \cos\theta_{\L}^{\rm m}(T)  - N_{\rm S}\,\sin\theta_{\L}^{\rm m}(T)  \\
N_{\Romannum{2}}^{\rm m}(T)  & = & N_{\rm DM} \, 
\sin\theta_{\L}^{\rm m}(T)  + N_{\rm S} \, \cos\theta_{\L}^{\rm m}(T) \,  ,
\nonumber
\eea
where $\theta^{\rm m}_{\L}$ is the mixing angle in matter describing the transformation from 
mass-Yukawa eigenstates to matter eigenstates given by
\be
\tan 2\theta_{\L}^{\rm m}= 
\frac{\sin 2\theta_{\L}}{1  + v^Y_{\rm S}}  \,  ,
\ee 
where, since $\theta_{\L} \lll 1$, we approximated $\cos \theta_{\L} \simeq 1$.
}
just a tiny fraction of $N_{\rm S}$'s, produced by Yukawa interactions, is converted into $N_{\rm DM}$'s at the  resonance.  
However, since we are considering heavy DM RH neutrinos, even a tiny amount can be sufficient
to reproduce the observed DM abundance.
Indeed, the relic DM abundance can be expressed in terms of the DM conversion fraction $(N_{N_{\rm DM}}/N_{N_{\rm S}})_{\rm res}$ 
at the resonance simply as\footnote{
For a generic DM scenario one can write \cite{book}
\be\label{ODMh2theory}
\O_{\rm DM}\,h^2 = {n_{\rm DM 0}\,M_{\rm DM} \over \ve_{\rm c0}\,h^{-2}}
= {n_{\g 0} \, M_{\rm DM} \over \ve_{\rm c0}\,h^{-2}\,f(t_{\rm f},t_0)} \,  
\left({N_{N_{\rm DM}} \over N_{\g}}\right)_{\rm f} \simeq 
1.45 \times 10^6 \, \left({N_{N_{\rm DM}} \over N_{\g}}\right)_{\rm f} \, \left({M_{\rm DM} \over {\rm GeV}}\right) \,  ,
\ee
where $\ve_{{\rm c} 0} \simeq 10.54 \,h^2 \,{\rm GeV}\,{\rm m^{-3}}$,
$n_{{\rm DM} 0}$ and $n_{\gamma 0} \simeq 410.7 \times 10^{-6} \,{\rm m}^{-3}$ are 
respectively the critical energy density, the DM number density and  the relic photon number density at the present time. 
With  the subscript `f' we are indicating 
the DM abundance freezing time. We are also indicating with
$f(t_{\rm f},t_0) = N_{\g 0} / N_{\g}^{\rm f} = g_{S {\rm f}}/g_{S0} =106.75 \times 11 /43 \simeq 27.3$ 
the dilution factor between the freezing time and the present time calculated within a standard cosmological model
(entropy production is negligible \cite{pdb}).
Possible further dilution due to the same degrees of freedom that play a role in 
DM genesis, in our case the heavy RH neutrinos, 
is  included in the calculation of $\left({N_{N_{\rm DM}}/ N_{\g}}\right)_{\rm f}$.
This expression gets specialised in our case, within a LZ approach, assuming $t_{\rm f} = t_{\rm res}$ 
and that all the DM abundance is instantaneously produced at $t_{\rm res}$
via non-adiabatic conversions.  We will see in the next section how this changes when 
the relic abundance is calculated within a more realistic density matrix formalism.}
 \be\label{DMabgres}
\O_{\rm DM}\,h^2 \simeq 1.45 \times 10^6 \, 
\left(\frac{N_{N_{\rm S}}}{N_\g}\right)_{\rm res} \,
\left(\frac{N_{N_{\rm DM}}}{N_{N_{\rm S}}}\right)_{\rm res} \, \left(\frac{M_{\rm DM}}{{\rm GeV}}\right)   \,  ,
\ee
 where $(N_{N_{\rm S}}/N_\g)_{\rm res}$ is
 the source RH neutrino-to-photon  ratio at the resonance.
 This is an important parameter determined by the initial conditions.
The latest 2018 {\em Planck} satellite results find for the DM abundance at the present time
(combining temperature and polarization anisotropies and gravitational lensing)~\cite{planck18}
\be\label{measured}
\O_{\rm DM}\,h^2 = 0.11933 \pm 0.00091  \,  .
\ee
Therefore, one can see that one can reproduce the measured value for 
$(N_{N_{\rm DM}}/N_{N_{\rm S}})_{\rm res} \sim 10^{-10}\,({\rm TeV}/M_{\rm DM})/(N_{N_{\rm S}}/N_\g)_{\rm res}$,
indeed a tiny amount if $M_{\rm DM} \gtrsim 1\,{\rm TeV}$, as in our case. This is  a basic observation on which 
the mechanism relies.  

A simple way to calculate $(N_{N_{\rm DM}}/N_{N_{\rm S}})_{\rm res}$, 
\cite{ad,unified},  is given by the LZ formula,
\be\label{NDM2NS}
\left.\frac{N_{N_{\rm DM}}}{N_{N_{\rm S}}}\right|_{\rm res} \simeq  \frac{\pi}{2} \, \g_{\rm res}  \,  ,
\ee
where the {\em adiabaticity parameter at the resonance}, $\g_{\rm res}$, is defined as 
\be
\g_{\rm res}  \equiv  \left. \frac{|E^{\rm m}_{\rm DM}-E^{\rm m}_{\rm S}|}{2 \, |\dot{\theta}_m|} \right|_{\rm res}  \,  ,
\ee
and in our specific case one finds 
\be\label{gres0}
\g_{\rm res}  = \sin^2 2\theta_{\L}(T_{\rm res}) \, \frac{|\D M^2|}{12\,T_{\rm res}\,H_{\rm res}}  
\simeq 
0.4 \, \frac{M_{\rm Pl} \, \sqrt{|\D M^2|}}{\widetilde{\L}^2 \,\sqrt{g^{\rm res}_{\star}}\, h_{\rm S}^3}  \,  ,
\ee
where $g^{\rm res}_{\star}=g_{\star}^{SM} + g_{\star}^{N_S} = 
106.75 + g_{\star}^{N_S} \simeq 106.75$ is the number of ultra-relativistic degrees of freedom at the resonance
given basically by the SM value and in the second numerical equation we used for the expansion rate at the
resonance $H_{\rm res} \simeq 1.66\,\sqrt{g^{\rm res}_{\star}}\,T^2_{\rm res}/M_{\rm Pl}$. 

Let us now  take into account the constraints on
neutrino masses from the seesaw formula and neutrino mixing experimental results. 
To this extent, it is useful to define the effective neutrino mass associated to the source RH neutrino,
$\widetilde{m}_S \equiv v^2 \, h^2_S / M_S $.
This provides an easy way to normalise  the Yukawa couplings  
taking  automatically into account the seesaw formula and  the 
information on neutrino masses from neutrino mixing experiments. 
Indeed if we define $\a_S \equiv \widetilde{m}_S/m_{\rm sol}$,
where $m_{\rm sol}$ is the solar neutrino mass scale, then necessarily, from the seesaw formula, 
one has  $\a_S \geq 1$.  Notice that $h_{\rm S}$ can be conveniently expressed in terms of $\a_{\rm S}$ as
\be\label{h2S}
h^2_{\rm S} = \a_{\rm S} \, {m_{\rm sol}\, M_{\rm S} \over v^2} \,   .
\ee
Introducing the variable $z \equiv M_{\rm DM}/T$ and its value at the resonance, 
$z_{\rm res} \equiv M_{\rm DM}/T_{\rm res}$,
 one can express $\D M^2$ in terms of $z_{\rm res}$ finding
\be\label{DM2}
\sqrt{|\D M^2|} = \frac{h_{\rm S}\,M_{\rm DM}}{2 \, z_{\rm res}} ~.
\ee
Using this relation in Eq.~(\ref{gres0}) and using the definition of $\a_{\rm S}$, 
one can then conveniently express the adiabaticity parameter at the resonance as \cite{unified}
\be\label{gres2}
\g_{\rm res} \simeq \frac{8}{\a_{\rm S}\,z_{\rm res}}
\, \left(\frac{M_{\rm DM}}{M_{\rm S}}\right) \, \left(\frac{10^{16} \, {\rm GeV}}{\widetilde{\L}}\right)^2  \,  .
\ee
Plugging this expression into the LZ formula Eq.~(\ref{NDM2NS}), one obtains first
$(N_{N_{\rm DM}}/N_{N_{\rm S}})_{\rm res}$ and then, from  
Eq.~(\ref{DMabgres}), for the DM abundance \cite{unified}:
\be\label{DMab}
\O_{\rm DM}\,h^2 \simeq \frac{0.1822}{\a_{\rm S} \, z_{\rm res}} \, 
\left(\frac{N_{N_{\rm S}}}{N_\g}\right)_{\rm res} \,
\left(\frac{M_{\rm DM}}{{\rm GeV}}\right) \,
\left(\frac{M_{\rm DM}}{M_{\rm S}}\right) \,
\left(\frac{10^{20} \, {\rm GeV}}{\widetilde{\L}}\right)^2 \,  \,  .
\ee
Imposing that this expression reproduces the measured value Eq.~(\ref{measured}),
one obtains the value of $\widetilde{\L}$ that reproduces the observed DM abundance 
\be\label{LAMBDADM}
\widetilde{\L}_{\rm DM} \simeq 10^{20} \, {\rm GeV} \,\sqrt{\frac{1.53}{\a_{\rm S} \, z_{\rm res}} \, 
\left(\frac{N_{N_{\rm S}}}{N_\g}\right)_{\rm res} \,\frac{M_{\rm DM}}{M_{\rm S}}\, \frac{M_{\rm DM}}{{\rm GeV}}}  \, \,  .
\ee 
In this expression one can see that there are five parameters: $z_{\rm res}$, $\a_{\rm S}$,  
$(N_{N_{\rm S}}/N_\g)_{\rm res}$, $M_{\rm DM}$ and $M_{\rm DM}/M_{\rm S}$ or, alternatively, $M_{\rm S}$.
However, from the relation (\ref{DM2}), $z_{\rm res}$ can actually
be expressed in terms of $\a_{\rm S}$, $M_{\rm DM}$ and $M_{\rm DM}/M_{\rm S}$ as 
\be\label{zres}
z_{\rm res}  = \frac{h_{\rm S} \, M_{\rm DM}}{2 \, \sqrt{M^2_{\rm DM}-M^2_{\rm S}}} \simeq  0.85 \times 10^{-8}\,\sqrt{\a_{\rm S}\,{M_{\rm S}\over M_{\rm DM}} \,\left({M_{\rm DM}\over {\rm GeV}}\right)}
\frac{M_{\rm DM}/M_{\rm S}}{\sqrt{M^2_{\rm DM}/M^2_{\rm S}-1}}  \,  ,
\ee
showing that there are actually only four independent parameters. This expression for $z_{\rm res}$ 
also shows that $z_{\rm res} \ll 1$, or equivalently $T_{\rm res} \gg M_{\rm DM}$, 
implying that  the reheat temperature  $T_{\rm RH} > T_{\rm res}$ cannot be too low
within this description.\footnote{The condition that $T_{\rm RH}$ needs to be higher than $T_{\rm res}$ 
comes from the fact that at the resonance the Higgs must thermalise in order to produce medium effects 
via the effective potential (\ref{LAMBDAmixing}) generated by the Anisimov operator.}
Since there is an upper bound $T_{\rm RH} \lesssim 10^{15}\,{\rm GeV}$,
this implies some  constraints on the allowed region of parameters. 

The dependence on the initial conditions is encoded in the value  $(N_{N_{\rm S}}/N_\g)_{\rm res}$.
If one assumes that some mechanism is able to thermalise the source RH neutrinos prior to the
resonant conversion, then $(N_{N_{\rm S}}/N_\g)_{\rm res} =3/4$, and in this case one obtains
\be\label{LAMBDADMthermal}
\widetilde{\L}_{\rm DM} \simeq 10^{20} \, {\rm GeV} \,\sqrt{\frac{1.15}{\a_{\rm S} \, z_{\rm res}} \, 
\frac{M_{\rm DM}}{M_{\rm S}}\, \frac{M_{\rm DM}}{{\rm GeV}}}  \, \,  .
\ee 
A more interesting case, since it does not rely on any external mechanism, is to assume that,
after inflation, the $N_{\rm S}$-abundance vanishes and is then produced
by the thermal bath through the Yukawa interactions.  The production is described by the simple
rate equation 
\be\label{rateequation}
\frac{dN_{N_{\rm S}}}{dz_{\rm S}} = - (D_{\rm S}+S_{\rm S})\,(N_{N_{\rm S}}-N_{N_{\rm S}}^{\rm eq})  \,  ,
\ee
where we defined $z_{\rm S} \equiv M_{\rm S}/T = z\,M_{\rm S} /M_{\rm DM}$, 
$D_{\rm S} \equiv \G^{\rm S}_D/(H\,z_{\rm S})$, $S_{\rm S} \equiv \G^{\rm S}_{\rm S}/(H\,z_{\rm S})$ and indicated
with   $\G^{\rm S}_{\rm D}$  and $\G^{\rm S}_{S}$ the source RH neutrino total decay  and
$\D L = 1$ scattering rates respectively and with $H$  the expansion rate. 
Moreover, we are normalising the abundances in  way that the thermal equilibrium $N_{\rm S}$-abundance is given by
\be\label{NNeq}
N_{N_{\rm S}}^{\rm eq}(z_{\rm S}) =  
{1\over 2}\,\int_{z_{\rm S}}^{\infty}dx\,\,x\,\sqrt{x^2-z_{\rm S}^2}\,e^{-x}\,  .
\ee 
In particular, in the ultra-relativistic equilibrium, one has $N_{N_{\rm S}}^{\rm eq}(z_{\rm S}\ll 1) =1$.
Since we are now assuming initial vanishing $N_{\rm S}$-abundance, until decays are negligible compared to inverse decays and 
$N_{N_{\rm S}} \ll N_{N_{\rm S}}^{\rm eq}$,  the rate equation is approximated by 
\be\label{rateequationsimple}
\frac{dN_{N_{\rm S}}}{dz_{\rm S}} \simeq (D_{\rm S}+S_{\rm S})\, N_{N_{\rm S}}^{\rm eq} \,  .
\ee
Since $(D_{\rm S}+S_{\rm S})(z_{\rm S}\ll 1) \simeq K_{\rm S}/5 = 1/z_{\rm S}^{\rm eq}$, 
where $z_{\rm S}^{\rm eq}\simeq 0.5\,\a_{\rm S}^{-1}$ \cite{garbrecht,unified}
is the value of $z_{\rm S}$ at the time when the $N_{\rm S}$-abundance thermalises,
one obtains the simple solution
\be
N_{N_{\rm S}}(z < z_{\rm S}^{\rm eq}) \simeq \frac{z_{\rm S}}{z_{\rm S}^{\rm eq}}   
\ee
and at the resonance one has \cite{unified}
\be\label{NSzreshier}
\left(\frac{N_{N_{\rm S}}}{N_\g}\right)_{\rm res} \simeq
{3\over 4}\,{z_{\rm res}\over z_{\rm S}^{\rm eq}}\,{M_{\rm S}\over M_{\rm DM}} \,  .
\ee 
Consequently, one obtains for the scale of new physics reproducing the observed DM abundance
\be\label{LAMBDADMvanishing}
\widetilde{\L}_{\rm DM} \simeq 10^{20} \, {\rm GeV} \,\sqrt{\frac{1.15}{\a_{\rm S} \, z_{\rm S}^{\rm eq}} \, 
\frac{M_{\rm DM}}{{\rm GeV}}}  \, \,  ,
\ee 
showing that it is indeed convenient having expressed $\widetilde{\L}_{\rm DM}$
in terms of $z_{\rm res}$ in the general relation (\ref{LAMBDADM}),
since in this way it cancelled out. 

These results show that Higgs induced interactions 
in Eq.~(\ref{anisimov}) are potentially able to reproduce the correct DM abundance for a proper
choice of parameters.
However, the same Higgs induced interactions are also responsible for  the $N_{\rm DM}$'s to decay at the present time, something that implies both constraints to be imposed but also an opportunity to test the scenario, 
in particular by studying the very high energy neutrino flux discovered at IceCube.\footnote{Plus of course in the next section we need to test how these results change solving the density matrix equation.}

There are two  decay channels to be taken into account. The first
one is the two body decay process $N_{\rm DM} \rightarrow A + \ell_{\rm S}$, where $A$ is a gauge boson
and ${\ell}_{\rm S}$ is either a charged lepton or a neutrino with a flavour composition determined by the
$N_{\rm S}$ Yukawa couplings \cite{unified}.
This occurs because even at zero temperature, after electroweak spontaneous symmetry breaking, the Anisimov operator
still generates a small (vacuum) mixing angle between $N_{\rm DM}$ and $N_{\rm S}$ given by\footnote{This expression
contains a factor 2 that was missed in \cite{unified} and that of course goes in the direction to
make bounds more stringent. See Appendix A for details on the derivation.}
\be\label{thetaL0}
\theta_{\L 0} = 
\frac{2\, v^2/\widetilde{\L}}{M_{\rm DM}\,(1-M_{\rm S}/M_{\rm DM})} \,  .
\ee
This mixing results in the two body decay of $N_{\rm DM}$ with rate\footnote{ In the absence
of the Higgs induced interactions, $N_{\rm DM}$ and $N_{\rm S}$ 
would coincide with the energy eigenstates and $N_{\rm DM}$
would be stable. However, when Higgs induced interactions are turned on, they generate a small 
non-diagonal Majorana mass term that  breaks the symmetry responsible
for the vanishing of the DM RH neutrino Yukawa couplings and its stability (see Appendix A).  
Indeed, if one considers  a $Z_2$ symmetry, the Anisimov operator is not invariant under this symmetry.
Notice that this expression is proportional to $M_{\rm DM}$, correcting the one given
in \cite{unified} (proportional to $M_{\rm S}$).}
\be
\Gamma_{{\rm DM}  \ra A + \ell_{\rm S}}  =    
\theta_{\L 0}^{2} \, {h^2_{\rm S} \over 4\, \pi}\, \,M_{\rm DM} \,  .
\ee
Inserting the expression for $\theta_{\L 0}$, 
one then obtains for the inverse decay rate
\be
\Gamma^{-1}_{{\rm DM} \ra A + \ell_{\rm S}}  = 
{\pi \over h^2_{\rm S}} \, \left(\frac{\widetilde{\L}}{v^2}\right)^2
\, M_{\rm DM}\,\left(1-{M_{\rm S}\over M_{\rm DM}}\right)^2\,  .
\ee

Using Eq.~(\ref{h2S}) and imposing  $\widetilde{\L}=\widetilde{\L}_{\rm DM}$, with $\widetilde{\L}_{\rm DM}$
given by Eq.~(\ref{LAMBDADM}), one then finds\footnote{Notice that
${\rm GeV}^{-1} \simeq 6.7 \times 10^{-25} \,{\rm s}$.}
\be
\G^{-1}_{{\rm DM} \ra A + \ell_{\rm S}}  \simeq
1.25 \times {10^{23}\,{\rm s} \over \a_{\rm S}^2\,z_{\rm res}}\,
\left(\frac{N_{N_{\rm S}}}{N_\g}\right)_{\rm res} \, {M_{\rm DM} \over {\rm GeV}} \,
\left({M_{\rm DM}\over M_{\rm S}}\right)^2 \,
\left(1-{M_{\rm S}\over M_{\rm DM}}\right)^2 \,  .
\ee
IceCube data constrains the lifetime to be longer than $\tau_{\rm DM}^{\rm min} \sim 10^{28}\,{\rm s}$, since otherwise 
an associated high energy neutrino flux would have been observed.  Therefore, imposing
$\G^{-1}_{{\rm DM} \ra A + \ell_{\rm S}} \geq \t_{\rm DM}^{\rm min}$, one obtains a lower bound on $M_{\rm DM}$.

For {\em initial thermal $N_{\rm S}$-abundance} one has $(N_{N_{\rm S}}/N_{\g})_{\rm res} = 3/4$ and 
using the expression (\ref{zres}) for $z_{\rm res}$ one obtains a lower bound that is much below
the Higgs mass and that is, therefore, meaningless since we are assuming  $N_{\rm DM}$ to be
heavier than the Higgs boson. 

For {\em initial vanishing $N_{\rm S}$-abundance} one can use Eq.~(\ref{NSzreshier})
for $(N_{N_{\rm S}}/N_{\g})_{\rm res}$ and in this case one obtains, in the hierarchical case 
$M_{\rm DM} \gg M_{\rm S}$, the lower bound
\be\label{lowerboundvan}
M_{\rm DM} \geq M_{\rm DM}^{\rm min} \simeq 54 \, {\rm TeV} \, 
\a_{\rm S}\,\t_{28}\,\left({M_{\rm S}\over M_{\rm DM}}\right) \,  ,
\ee
where we defined $\tau_{28} \equiv \tau_{\rm DM}^{\rm min}/(10^{28}\,{\rm s})$. 

Another important decay channel for $N_{\rm DM}$ at the present time is the four body
decay $N_{\rm DM} \ra 3\,A + \ell_{\rm S}$. In the narrow width approximation the decay
rate is given by \cite{unified}
\be
\G_{{\rm DM} \ra 3A + {\ell}_{\rm S}} = 
{\Gamma_{\rm S} \over 15 \cdot 2^{11} \cdot \pi^{4}} \, {M_{\rm DM} \over M_{\rm S}} \, 
\left({M_{\rm DM} \over \widetilde{\L}}\right)^2 \,  ,
\ee
where $\G_{\rm S} = h^2_{\rm S}\,M_{\rm S}/(4\,\pi)$. 
It is important to notice that this expression is valid for $M_{\rm S} > M_{W} \sim 100 \,{\rm GeV}$.
For lower masses the source RH neutrino decays can occur via three  body decays, corresponding to
five body decays for $N_{\rm DM}$, and
the decay rate is greatly suppressed and does not produce significant constraints.\footnote{Indeed
for three body decays, the cross section is phase space suppressed by the fifth power of the mass
of the decaying particle. Notice that this case, for $M_{\rm S} < M_W$, has not been considered in \cite{unified}.} 

Using again Eq.~(\ref{h2S})
to express $h^2_{\rm S}$ in terms of $\a_{\rm S}$ and imposing $\widetilde{\L} = \widetilde{\L}_{\rm DM}$
(see Eq.~(\ref{LAMBDADM})) one finds for the inverse decay rate
\be\label{tau4b}
\G^{-1}_{{\rm DM} \ra 3A + {\ell}_{\rm S}} \simeq 
{0.153 \times 10^{40} \,{\rm s} \over \a_{\rm S}\,z_{\rm res}} \,
\left(N_{N_{\rm S}}\over N_{\g}\right)_{\rm res} \,
\left({M_{\rm DM} \over M_{\rm S}}\right)^2 \,  
\left({{\rm GeV} \over M_{\rm DM}}\right)^3  \,  .
\ee
Imposing again that the lifetime is sufficiently long to escape IceCube constraints implies
 $\G^{-1}_{{\rm DM} \ra 3A + {\ell}_{\rm S}} \geq 10^{28}\,{\rm s}\,\tau_{28}$ that
 this time  leads to an upper bound on the DM RH neutrino mass given by
 \be\label{upperbound}
 M_{\rm DM} \lesssim 5.3 \, {\rm TeV} \, \a_{\rm S}^{-{2\over 3}} 
 \, z_{\rm res}^{-{1\over 3}} \, \tau_{28}^{-{1\over 3}}\,
 \left(N_{N_{\rm S}}\over N_{\g}\right)_{\rm res}^{1\over 3} \,
 \left({M_{\rm DM} \over M_{\rm S}}\right)^{2\over 3} \,  .
 \ee
We can again specialise this upper bound first to the case of {\em initial  thermal
$N_{\rm S}$-abundance}, for $\left(N_{N_{\rm S}}/ N_{\g}\right)_{\rm res} = 3/4$, finding 
\be
M_{\rm DM} \lesssim 4.8 \, {\rm TeV} \, \a_{\rm S}^{-{2\over 3}} 
 \, z_{\rm res}^{-{1\over 3}} \, \tau_{28}^{-{1\over 3}}\,
 \left({M_{\rm DM} \over M_{\rm S}}\right)^{2\over 3} \,   ,
\ee
and from this, using Eq.~(\ref{zres}) for $z_{\rm res}$, one finds
\be\label{upperboundthermal}
M_{\rm DM} \lesssim 0.3 \,{\rm PeV} \, \a_{\rm S}^{-{5\over 7}} \, \t_{28}^{-{2\over 7}}
\,\left({M_{\rm DM} \over M_{\rm S}} \right)^{5\over 7} \, 
\, \left[\frac{M_{\rm DM}/M_{\rm S}}{\sqrt{M^2_{\rm DM}/M^2_{\rm S}-1}}\right]^{-{2\over 7}} \,  .
\ee
One can notice  again that the most conservative bound is  obtained for $\a_{\rm S}=1$, 
since this minimises the $N_{\rm S}$ Yukawa coupling making $N_{\rm DM}$ more stable.
Moreover again higher values of $M_{\rm DM}/M_{\rm S}$ tend to relax also this upper bound.

However, there is an additional constraint coming from the requirement
$T_{\rm res} \leq T_{\rm RH} \lesssim 10^{15} \, {\rm GeV}$
translating into an upper bound on $M_{\rm DM}/M_{\rm S}$
that can be derived combining Eq.~(\ref{zres}) for $z_{\rm res}$ with 
Eq.~(\ref{upperboundthermal}) for $M_{\rm DM}^{\rm max}$, obtaining
\be\label{upperboundmassratiothermal}
{M_{\rm DM} \over M_{\rm S}} \lesssim 8 \times 10^4 \,\a_{\rm S} \, \t_{28}^{1\over 6} \,  ,
\ee
corresponding to an absolute  upper bound on the DM mass\footnote{Notice that the bound is saturated for
$M_{\rm S} = 10^4\,{\rm GeV}\,\a_{\rm S}^{-1}\,\tau_{28}^{-{1\over 3}}$, so that the 
assumption $M_{\rm S} > M_{W}$ for the four body decays decay constraints holds.}
\be\label{absub}
M_{\rm DM} \lesssim 1.0 \times 10^9 \, {\rm GeV} \, \tau_{28}^{-{1\over 6}}  \,  .
\ee
Considering the {\em case of initial vanishing $N_{\rm S}$-abundance}, substituting
Eq.~(\ref{NSzreshier}) for $\left(N_{N_{\rm S}}/ N_{\g}\right)_{\rm res}$ into 
Eq.~(\ref{upperbound}), one finds \cite{unified}
\be\label{upperboundvan}
M_{\rm DM} \leq M_{\rm DM}^{\rm max} \simeq 6 \, {\rm TeV} \, \a_{\rm S}^{-{1\over 3}} \, \tau_{28}^{-{1\over 3}}\,
 \left({M_{\rm DM} \over M_{\rm S}}\right)^{1\over 3} \,   .
\ee
This upper bound combined with the lower bound (\ref{lowerboundvan}) identifies
an allowed window for the value of the DM mass that, however, because of the
lower bound on the lifetime $\tau \geq \tau_{\rm DM}^{\rm min}$, opens up only in the hierarchical case,
for sufficiently large $M_{\rm DM}/M_{\rm S}$. Imposing 
\be\label{lifetimebound}
\tau \simeq (\G_{{\rm DM} \ra A + \ell_{\rm S}}  + \G_{{\rm DM} \ra 3A + {\ell}_{\rm S}})^{-1} > 
\tau_{\rm DM}^{\rm min} \simeq 10^{28} \, {\rm s} \,  , 
\ee
 one finds\footnote{This result is clearly more stringent than the result 
$M_{\rm DM}/M_{\rm S} \gtrsim 2.3\,\a_{\rm S} \, \tau_{28}$
found in \cite{unified} because of the more stringent  lower bound on $M_{\rm DM}$
as an effect of the two corrections we found to the rate $\G_{{\rm DM} \ra A + {\ell_{\rm S}}}$
and also because we are more accurately taking the inverse of the sum of the rates to calculate the life time
in the regime where the two rates are comparable.}
$M_{\rm DM}/M_{\rm S} \gtrsim 10 \, \a_{\rm S} \, \tau_{28}$, with 
the allowed region opening up when the lower bound on $M_{\rm DM}/M_{\rm S}$ saturates 
at a  value  $M_{\rm DM}^\star \simeq 8\,{\rm TeV}$.

Notice that since the upper bound (\ref{upperboundvan}) applies only for $M_{\rm S} > M_W \sim 100\,{\rm GeV}$, then one has to impose $M_{\rm DM}/M_{\rm S} \lesssim 10^{3}\,M_{\rm DM}/{\rm TeV}$.  
This implies that for masses $M_{\rm S} > M_W$ the upper bound can be relaxed only up to
$M_{\rm DM} \lesssim 150\,{\rm TeV}\,\a_{\rm S}^{-{1\over 2}}\,\tau_{28}^{-{1 \over 2}}$.
 
In Fig.~1 we show in purple, for the most conservative case $\a_{\rm S} = 1$, 
the allowed range on $M_{\rm DM}$ for $M_{\rm S} > M_W \sim 100\,{\rm GeV}$,
calculated, more accurately, using Eq.~(\ref{lifetimebound}) that also 
accounts for the lower bound (\ref{lowerboundvan}) from two body decays (this, however, holds also 
for $M_{\rm S} < M_{W}$).
In the case of initial vanishing $N_{\rm S}$-abundance,
the constraint $T_{\rm res} < T_{\rm RH} \lesssim 10^{15}\,{\rm GeV}$
is automatically satisfied in the region $M_{\rm S} > M_W \sim 100\,{\rm GeV}$.
For $M_{\rm S} < M_W$, as discussed, the upper bound on $M_{\rm DM}$ 
from four body decays does not apply and one is left only with the lower bound from
two body decays Eq.~(\ref{lowerboundvan}). However, there is still an upper bound on the reheat temperature
$T_{\rm res} < T_{\rm RH} < 10^{15}\,{\rm GeV}$, from Eq.~(\ref{zres}), that implies 
\be\label{ubTRHvan}
M_{\rm DM} \lesssim 0.85 \times 10^7 \, {\rm GeV}\,\sqrt{\a_{\rm S}\,{M_{\rm S}\over {\rm GeV}}} \,  .
\ee
In Fig.~1 we also show in orange the allowed range on $M_{\rm DM}$ for $M_{\rm S} > 1\,{\rm GeV}$
that is obtained combining the two bounds. 



Finally, one can impose {\em constraints from leptogenesis} \cite{fy}. 
As we have seen, within the scenario we discussed with $M_{\rm S} < M_{\rm DM}$, 
there is quite a stringent upper bound $M_{\rm DM} \lesssim 10^9\,{\rm GeV}$ (see Eq.~(\ref{absub})).
Moreover, the matter-antimatter asymmetry  has to be necessarily generated  from the decays of the source RH 
neutrinos, interfering with the third RH neutrino species in order to have non-vanishing $C\!P$ asymmetries \cite{unified}. 
Since $M_{\rm S} \ll M_{\rm DM} \lesssim 10^9\,{\rm GeV}$,  below the lower bound 
for successful leptogenesis in the two RH neutrino hierarchical case,  
$ M_{\rm lep }\lesssim 10^{10}\,{\rm GeV}$ \cite{lowerboundlep}, 
the source and the interfering RH neutrinos have to be necessarily quasi-degenerate 
in order to have sizeable $C\!P$ asymmetries resonantly enhanced \cite{crv}.  
Moreover, in order to have successful leptogenesis, the scale of generation of the asymmetry
has to be necessarily above the temperature at which sphaleron processes, converting part of the lepton asymmetry
into a baryon asymmetry, go out-of-equilibrium, with $T^{\rm off}_{\rm sph} \simeq 132\,{\rm GeV}$ \cite{sphaleron}.
Since in leptogenesis from decays the asymmetry is generated at a temperature that 
is at most half of the decaying RH neutrino mass, this requirement implies a lower bound \cite{unified}
$M_{\rm S} \gtrsim 300\,{\rm GeV}$, that can be also 
recast as a lower bound $M_{\rm DM}/M_{\rm S} \leq 3.3 \times 10^{-3}\,M_{\rm DM}/{\rm GeV}$.  
This upper bound on $M_{\rm DM}/M_{\rm S}$ can be easily combined with the bound on the DM lifetime
Eq.~(\ref{lifetimebound}).

In the case of initial thermal $N_{\rm S}$-abundance, it is easy to see that this lower bound
on $M_{\rm S}$, combined with the upper bound (\ref{upperboundthermal}),
leads to an upper bound on $M_{\rm DM}$ that is much less stringent
than the one (see Eq.~(\ref{absub})) coming from the upper bound on $T_{\rm RH}$.   

On the other hand, in the case of initial vanishing 
$N_{\rm S}$-abundance, one finds the approximate allowed region 
\be
4\,{\rm TeV}\,\a_{\rm S}^{{1\over 2}} \,\tau_{28}^{1\over 2} \lesssim M_{\rm DM} \lesssim 27 \, {\rm TeV} \,
\a_{\rm S}^{-{1\over 2}} \,\tau_{28}^{-{1\over 2}} \,  .
\ee
This is clearly more stringent both than the upper bound we derived for $M_{\rm S} > M_W$
and for $M_{\rm S} < M_W$, in this second case from the upper bound on the reheat temperature. 
This allowed region for successful leptogenesis, calculated more precisely from Eq.~(\ref{lifetimebound}),
is shown in green in Fig.~1 for the most conservative case $\a_{\rm S} = 1$ 
and one can notice that it is quite restricted.\footnote{It is interesting to notice that 
 if one would consider leptogenesis from RH neutrino oscillations, the so-called ARS scenario \cite{ARS},
 then since the asymmetry is produced when the source RH neutrino is ultra-relativistic, 
 the source RH neutrino mass can be much lighter and this would certainly highly 
 relax the constraint. A dedicated analysis would be certainly interesting in this respect.} In particular, one can notice
 that in this case there is quite a stringent upper bound on the DM lifetime
 $\tau_{\rm DM} \lesssim 4 \times 10^{28}\,{\rm s}$.
\begin{figure}
	\begin{center}
		\psfig{file=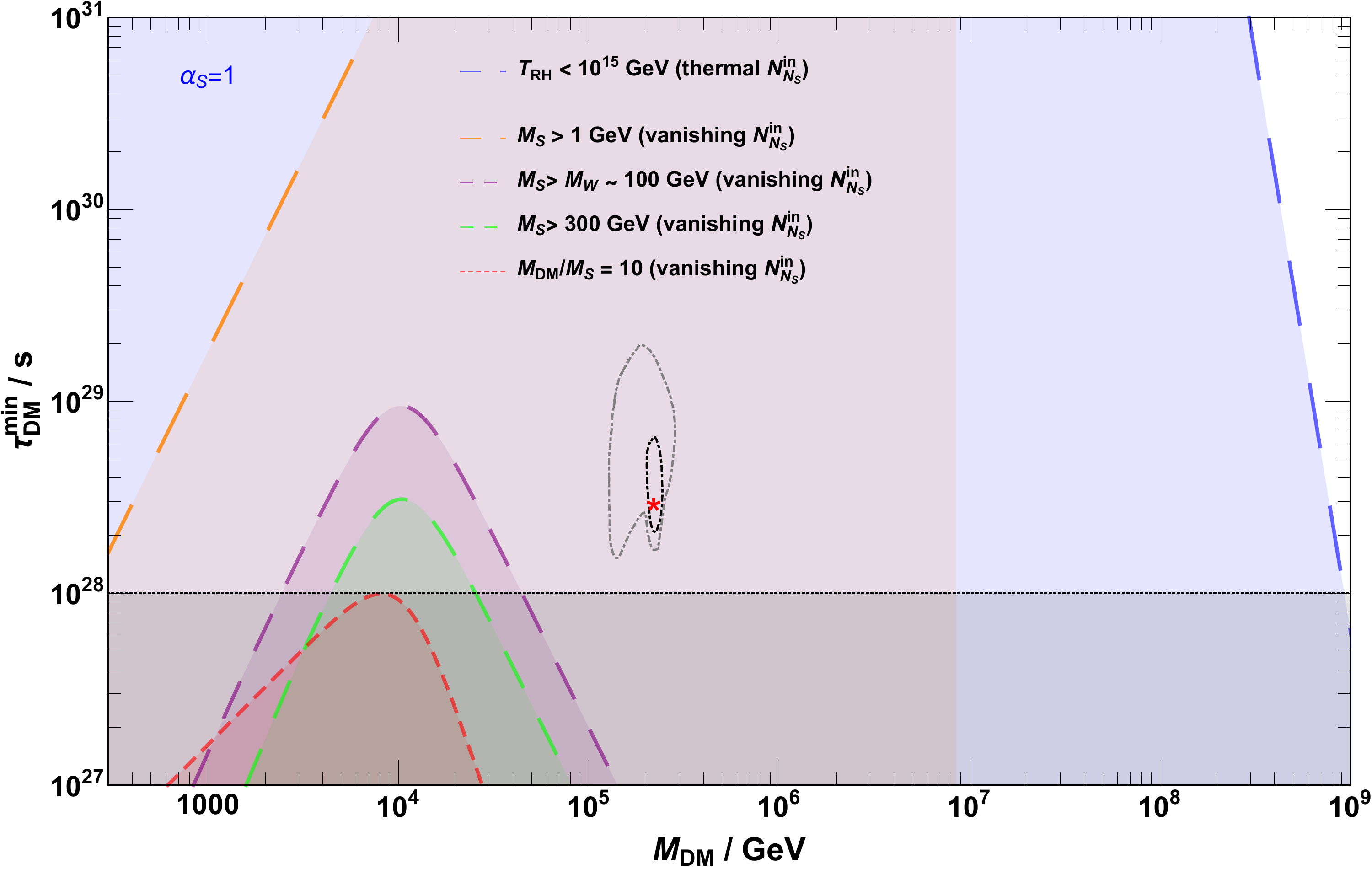,height=100mm,width=145mm}	\end{center}
	\vspace{-2mm}
	\caption{Summary of the allowed regions in the $M_{\rm DM}-\tau_{\rm DM}$ plane
	obtained within the LZ approximation for $\a_{\rm S}=1$ imposing different requirements.
	The region below the dotted black horizontal line, for $\tau_{\rm DM} < 10^{28}\,{\rm s}$, 
	is currently excluded by IceCube. 
	The light blue region delimited by the long dashed line is the region satisfying  
	$T_{\rm RH} < 10^{15} \, {\rm GeV}$ 
	for initial thermal  $N_{\rm S}$-abundance. 
	The orange region is for $M_{\rm S} > 1\,{\rm GeV}$ and vanishing $N_{N_{\rm S}}^{\rm in}$.  
	The vertical line corresponds to  the upper bound on $M_{\rm DM}$ Eq.~(\ref{ubTRHvan}) 
	from $T_{\rm RH} < 10^{15}\,{\rm GeV}$. 
	The green region satisfies
	the lower bound $M_{\rm S} > 300\,{\rm GeV}$ allowing also for successful leptogenesis.
	The red star, the black dotted line and the gray dotted line are respectively the best fit, $68\%$ and $95\%$ contour lines
	recently found in \cite{chianese3} analysing latest IceCube data including a contribution from DM (neutrinophilic) 
	decays in addition to an astrophysical component with Fermi spectrum.}
	\label{fig:lzboundsth}
\end{figure}
The existence of this upper bound shows that the possibility to combine DM with leptogenesis 
within this model will be certainly tested in the next future at neutrino telescopes. 
However, such a marginal allowed region legitimately questions whether a calculation of the DM abundance
within the simple LZ approximation gives the correct results, thus motivating a calculation
within a density matrix formalism.


\section{Density matrix formalism}

In this section we go beyond the LZ approximation and calculate the DM relic abundance 
within the density matrix formalism \cite{feynman}. 
The use of density matrix in neutrino physics in the early universe has a long history. The most traditional
 application is the study of active-sterile neutrino mixing in the early universe \cite{densitymearly}. 
In that case a comparison between the LZ approximation
and the density matrix formalism was made in \cite{foot} finding quite a good agreement.
The use of a density matrix formalism plays also a crucial role in the study of RH neutrino mixing 
in leptogenesis from neutrino oscillations \cite{ARS,arsdm}.
The density matrix formalism also proves to be very important  in the description of 
flavour effects in leptogenesis \cite{lepdm}. 

In the absence of Higgs induced interactions, the only interactions able to produce the source RH
neutrinos would be the Yukawa interactions so that the $N_{\rm DM}$'s would be completely 
decoupled. Therefore, Yukawa interactions would produce only source RH neutrinos 
(barring the third RH neutrino species for the time being).
This production can be described by a density matrix normalised in terms of the source RH neutrino abundance
that in the Yukawa basis would be diagonal and simply given by ($I,J=  {\rm DM}, {\rm S}$)
\be
N_{IJ}(z) = N_{N_{\rm S}}(z)\,\left(\begin{array}{cc}
0 & 0 \\
0 & 1
\end{array}\right) \,  .
\ee 
Here again we notice that we describe the system within a monochromatic approximation where 
momentum dependence is integrated away.  As we have seen, the abundance of source RH neutrinos, $N_{N_{\rm S}}$,
is described by the simple  rate equation (\ref{rateequation}).
However, when the Higgs induced interactions are turned on, they develop off-diagonal terms
that have to be taken into account together in principle with 
decoherence effects. This evolution is then described by a density matrix equation of the form \cite{densitymearly}
\be\label{densitymatrixeq}
{dN_{IJ} \over dt} = -i\,[{\cal H}, N]_{I J}  - 
\begin{pmatrix}
0   &  {1\over 2}(\G_D+\G_S) \,N_{\rm DM-S}  \\ 
{1\over 2}(\G_D+\G_S)  \,N_{\rm S-DM}  & (\G_D+\G_S)\,(N_{N_{\rm S}}-N_{N_{\rm S}}^{\rm eq})  
\end{pmatrix}
\,   ,
\ee
where the first term is the Liouville-von Neumann term and the second term is the combination 
of the decoherence term, damping off-diagonal terms, and  the repopulation (diagonal) term, describing the production 
of source RH neutrinos.

Clearly, without off-diagonal terms in the Hamiltonian, the density matrix equation would simply reduce  
to Eq.~(\ref{rateequation}). Moreover again a diagonal term in ${\cal H}$ cancels out and we can
replace ${\cal H} \ra \D{\cal H}$, with $\D{\cal H}$ given by Eq.~(\ref{effectiveham}).

As often done, we can express the matrices in the Pauli matrix basis using a vectorial notation.
The effective Hamiltonian in Eq.~(\ref{effectiveham}) can then be recast as
\be\label{DH}
\D{\cal H} = {1\over 2}\,\vec{V} \cdot \vec{\s}  \,  , 
\ee
where the {\em effective potential vector} $\vec{V}$ is defined as
\be
\vec{V} \equiv \frac{\D M^2}{6\,T}\,\left(\sin 2\theta_{\L} , 0 , - 1 - v_{\rm S}^Y \right)  \,   .
\ee 
The abundance density matrix is analogously recast, introducing the 
quantity $P_0$ and the polarisation vector $\vec{P}$, as \cite{volkas}
\be\label{N}
N =  {1\over 2}\,P_0 \, \left( 1 + \vec{P} \cdot \vec{\s} \right)\,  ,
\ee
in a way that 
\bea
N_{N_{\rm DM}}  &  =  &  {1\over 2}\,P_0 \, \left( 1 + P_z \right) \,   , \\ \nonumber 
N_{N_{\rm S}}     &  =  &  {1\over 2}\,P_0 \, \left( 1 - P_z \right) \,  ,\\  \nonumber
N_{N_{\rm DM}} + N_{N_{\rm S}} & = &  P_0 \,  . 
\eea
Inserting Eqs.~(\ref{DH}) and (\ref{N}) into the density matrix equation (\ref{densitymatrixeq}),
one obtains a set of equations for  $P_0$ and $\vec{P}$\footnote{In Appendix B we give some 
details on the derivation  and we also show a third alternative  equivalent way to 
write the density matrix equation often used in the literature.} 
\bea\label{densitymatrixeqpauli}
{d  \vec{P} \over d t}& = & \vec{V} \times \vec{P} -
\left[{1\over 2}(\G_D+\G_S)  + {d\ln P_0 \over dt} \right]\,\vec{P}_T - 
(1 + P_z)\,{d \ln P_0 \over d t} \,\hat{z} \,   , \\
{d  P_0 \over d t} & = & - (\G_D+\G_S)\,(N_{N_{\rm S}}-N_{N_{\rm S}}^{\rm eq})  \,   ,
\eea
where we defined $\vec{P}_T \equiv P_x\,\hat{x} + P_y \, \hat{y}$. 
If we explicitly unpack the first vectorial equation in terms of its components, 
we obtain the following set of four differential equations
\bea
{d  P_x \over d t}& = & - V_z \, P_y - {1\over 2}(\G_D+\G_S)  \,P_x - {P_x \over P_0}\,{dP_0 \over dt} \,  ,\\ \nonumber
{d  P_y \over d t}& = & V_z \, P_x -V_x \, P_z - {1\over 2}\,(\G_D+\G_S) \,P_y - {P_y \over P_0}\,{dP_0 \over dt} \, , \\ \nonumber
{d  P_z \over d t}& = & V_x \, P_y  - {1+P_z \over P_0}\,{dP_0 \over dt}  \, ,\\ \nonumber
{d  P_0 \over d t} & = & - (\G_D+\G_S)\,(N_{N_{\rm S}}-N_{N_{\rm S}}^{\rm eq})  \,   .
\eea
Changing the independent variable, from $t$ to  $z$, one obtains
\bea
{d  P_x \over d z}& = & - \overline{V}_z \, P_y 
- {1\over 2}\,{M_{\rm S} \over M_{\rm DM}} \,(D+S)  \,P_x - 
 {P_x \over P_0}\,{dP_0 \over dz} \,  , \\ \nonumber
{d  P_y \over d z}& = & \overline{V}_z \, P_x -\overline{V}_x \, P_z - {1\over 2}\,{M_{\rm S} \over M_{\rm DM}}\,(D+S) \,P_y -   {P_y \over P_0}\,{dP_0 \over dz} \,  , \\ \nonumber
{d  P_z \over d z}& = & \overline{V}_x \, P_y  - {1+P_z \over P_0}\,{dP_0 \over dz} \,  , \\ \nonumber
{d  P_0 \over d z} & = & - {M_{\rm S} \over M_{\rm DM}}\,(D+S)\,(N_{N_{\rm S}}-N_{N_{\rm S}}^{\rm eq})  \,   ,
\eea
where we have already defined $D$ and $S$ after Eq.~(\ref{rateequation}) and
we have now also introduced $\vec{\overline{V}} \equiv \vec{V}/(H\,z)$.

In the next section we show the evolution of the DM
abundance obtained solving numerically this set of density matrix equations
for a benchmark  value $M_{\rm DM} = 220\,{\rm TeV}$ and for different values
of $T_{RH}$, $\tau_{\rm DM}$ and $M_{\rm DM}/M_{\rm S}$. 

\section{Evolution of the DM abundance from the density matrix equation}

In this section we fix  the DM mass to a benchmark value $M_{\rm DM} = 220\,{\rm TeV}$ and we
show the evolution of the DM abundance, $N_{N_{\rm DM}}$, solving the density matrix equation
presented in the previous section. 
We choose this particular benchmark value for $M_{\rm DM}$ 
since it is the best fit value of DM mass found  in \cite{chianese3}, where the authors analysed 
IceCube data on the high energy neutrino flux energy spectrum within a model where, in addition to
an astrophysical component with a power-law spectrum with spectral index $\gamma = 2.2$, 
there is an additional contribution  from neutrinophilic DM decays.\footnote{The 
analysis does not straightforwardly translates to our model but it provides 
a good indication and motivation for  the use of such value of the DM mass as benchmark value.}

Though we fix $M_{\rm DM}$, we show how the evolution of the DM abundance depends on the other 
three parameters of the model: the reheat temperature $T_{\rm RH}$, 
the lifetime $\tau_{\rm DM}$ and finally the ratio $M_{\rm DM}/M_{\rm S}$
(or equivalently $M_{\rm S}$ considering that $M_{\rm DM}$ is fixed). 

In all plots we also show the relic value of the DM abundance, indicated with $N_{N_{\rm DM}}^{\rm f,obs}$, 
that  corresponds to the observed value for $\O_{\rm DM} h^2$ given in Eq.~(\ref{measured}).
This can be easily derived from Eq.~(\ref{ODMh2theory}), finding\footnote{Notice that with the 
normalisation we choose,
one has for the photon abundance $N_{\gamma}^{\rm f} = 4/3$, while for the photon
abundance at the present time one has $N_{\gamma 0} = N_{\gamma}^{\rm f}\,f(t_{\rm f},t_0) \simeq 36.4$.}
\be
N_{N_{\rm DM}}^{\rm f,obs} = (0.1097\pm0.0008) \times 10^{-6}\,\left({{\rm GeV}\over M_{\rm DM}}\right) \,  .
\ee
In particular, for our benchmark value $M_{\rm DM} =220 \,{\rm TeV}$, one finds 
$N_{N_{\rm DM}}^{\rm f,obs} \simeq 5 \times 10^{-13}$, the value indicated in Figs. 2--4 that we now briefly discuss.

\subsection{Dependence on the reheat temperature}

In Fig.~2 we show the dependence of $N_{N_{\rm DM}}(z)$ on the reheat temperature 
both for an initial thermal $N_{\rm S}$-abundance (upper panel)
and for an initial vanishing $N_{\rm S}$-abundance (lower panel).  In particular, we show 
$N_{N_{\rm DM}}(z)$ for different values of $T_{\rm RH}$ as indicated.
Notice that the  value $\tau_{\rm DM} = 3.46 \times 10^{28}\,{\rm s}$, corresponding to 
$\widetilde{\L} = 1.13 \times 10^{24}\,{\rm GeV}$, 
is just the value that reproduces the observed DM
abundance for $T_{\rm RH} =10^{15}\,{\rm GeV}$ and $M_{\rm S} = 300\,{\rm GeV}$
in the case of initial thermal $N_{\rm S}$-abundance.
In the case of initial vanishing $N_{\rm S}$-abundance this 
value for $\tau_{\rm DM}$ is too high, i.e., the coupling too small, 
to get the correct relic abundance even for maximum allowed $T_{\rm RH}$.
\begin{figure}
\begin{center}
\psfig{file=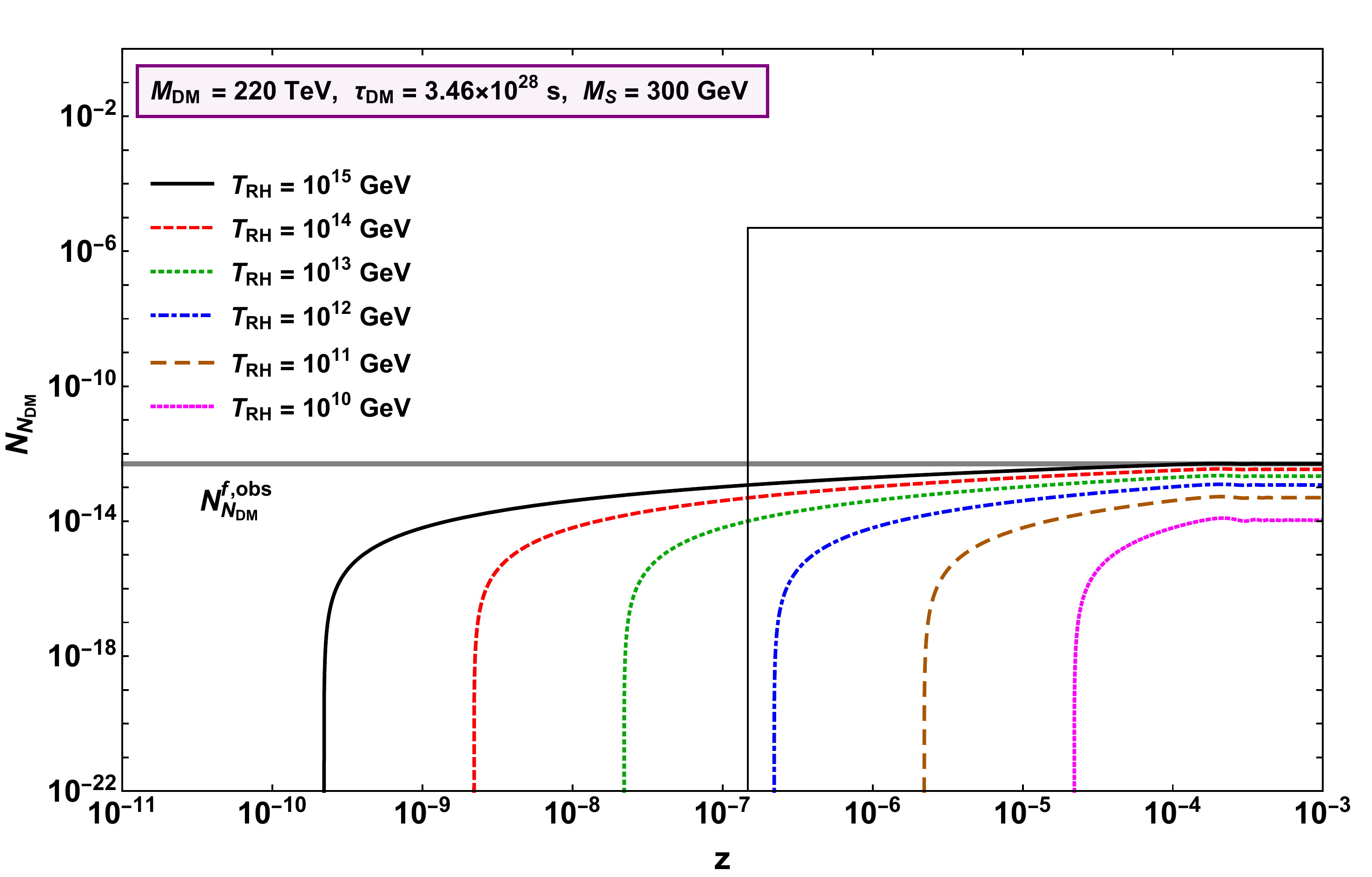,height=85mm,width=125mm} \\
\vspace{10mm}
\psfig{file=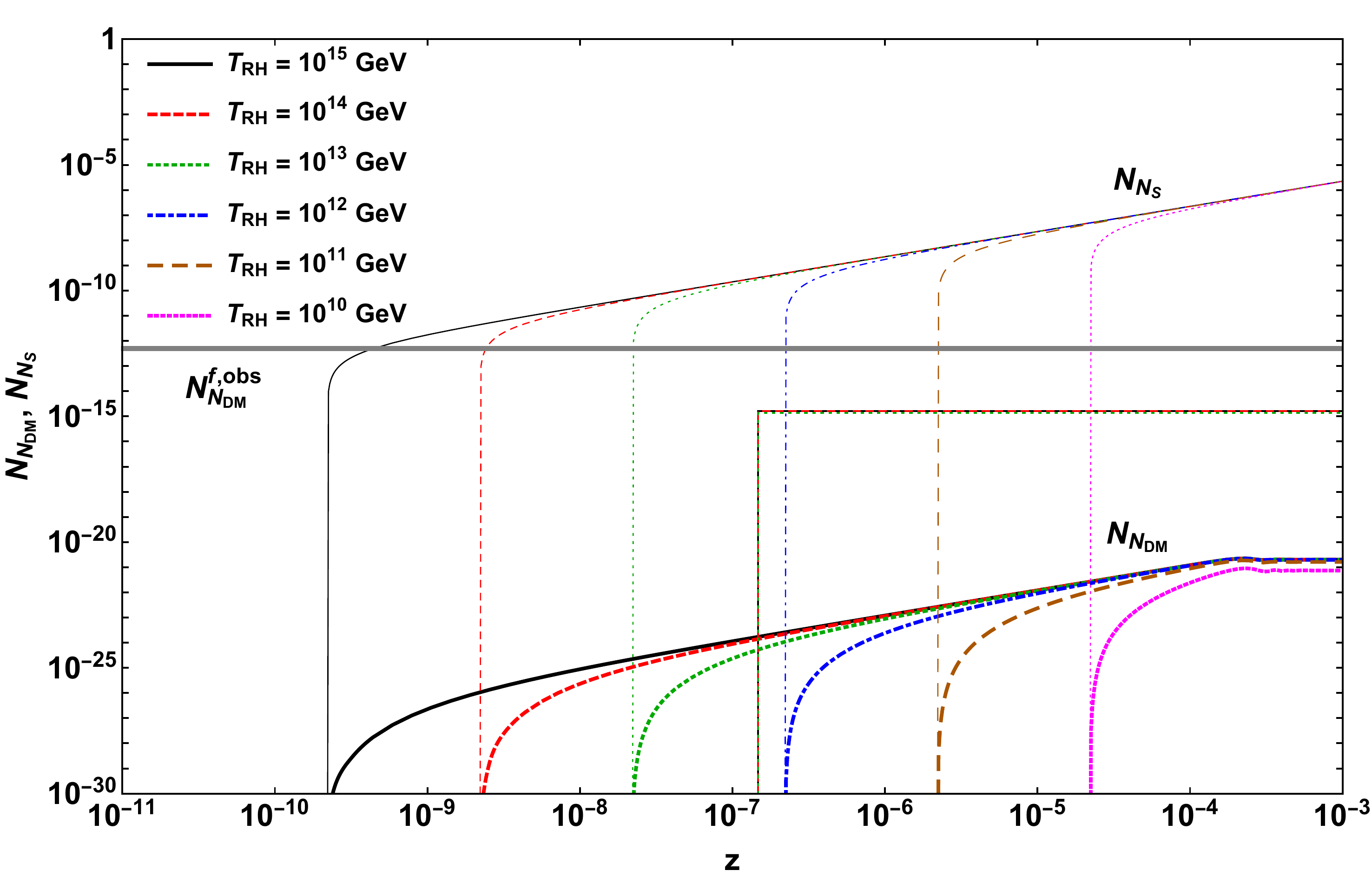,height=85mm,width=125mm}   
\end{center}
\vspace{-1mm}
\caption{Evolution of the DM abundance $N_{N_{\rm DM}}$ for different values
of $T_{\rm RH}$, as indicated, and for fixed values of $M_{\rm DM}, M_{\rm S}$ and $\tau_{\rm DM}$.
The upper (lower) panel is for initial thermal (vanishing) $N_{\rm S}$-abundance, i.e., $N^{\rm in}_{N_{\rm S}}=1$ in the upper panel. 
The step function corresponds to the instantaneous LZ approximation.}
\label{benchmark}
\end{figure}
It should be immediately noticed that the LZ approximation overestimates by many orders of magnitude the  relic DM abundance. 
It should also be noticed how in the case of initial vanishing $N_{\rm S}$-abundance the freeze-in temperature $T_{\rm f} \sim 10^9 \,{\rm GeV}$
is much below the resonant temperature $T_{\rm res} \sim 10^{12}\,{\rm GeV}$.\footnote{Using Eq.~(\ref{zres}) one finds
$z_{\rm res} \simeq 1.5 \times 10^7$ and since the decay rate in this case is dominated
by four body decays, from Eq.~(\ref{tau4b}) one finds $(\widetilde{\L}_{\rm DM}/\widetilde{\L})^2 \sim 10^7$,
translating into a final DM abundance in the case of LZ approximation that is seven orders of magnitude 
higher than the observed one, as it can be noticed in the plot.}
Another interesting thing to highlight is that for initial vanishing $N_{\rm S}$-abundance,
the relic value is basically independent of $T_{\rm RH}$, except for the lowest value $T_{\rm RH}=10^{10}\,{\rm GeV}$ 
when the production occurs close to the freezing and the relic value is not fully saturated.  
Therefore, $T_{\rm RH} \sim 10^{10}\,{\rm GeV}$ should be regarded as a border line value such that below this value the
production is strongly suppressed since there is no time for the asymmetry to be produced.  
These are all features that should be addressed by an analytical description. 

\subsection{Dependence on the lifetime}

In Fig.~3, we fix the reheat temperature to the highest possible value, $T_{\rm RH} = 10^{15}\,{\rm GeV}$,
and show how the evolution of the DM abundance depends on $\tau_{\rm DM}$.  Note how
for increasing values of $\tau_{\rm DM}$, corresponding to larger values of $\widetilde{\Lambda}$, the relic
DM abundance decreases and vice-versa. It can be noticed how in the case of initial vanishing  
$N_{\rm S}$-abundance, even for a very low (and excluded by experimental data) value
$\tau_{\rm DM} = 10^{24}\,{\rm s}$, corresponding to $\widetilde{\L}= 6.4 \times 10^{21}\,{\rm GeV}$, 
the relic DM abundance is a few orders of magnitude
below the measured value. Notice also how, though the final relic DM abundance is clearly strongly depending on $\tau_{\rm DM}$,
the freeze-in temperature is not. 
\begin{figure}
\begin{center}
\psfig{file=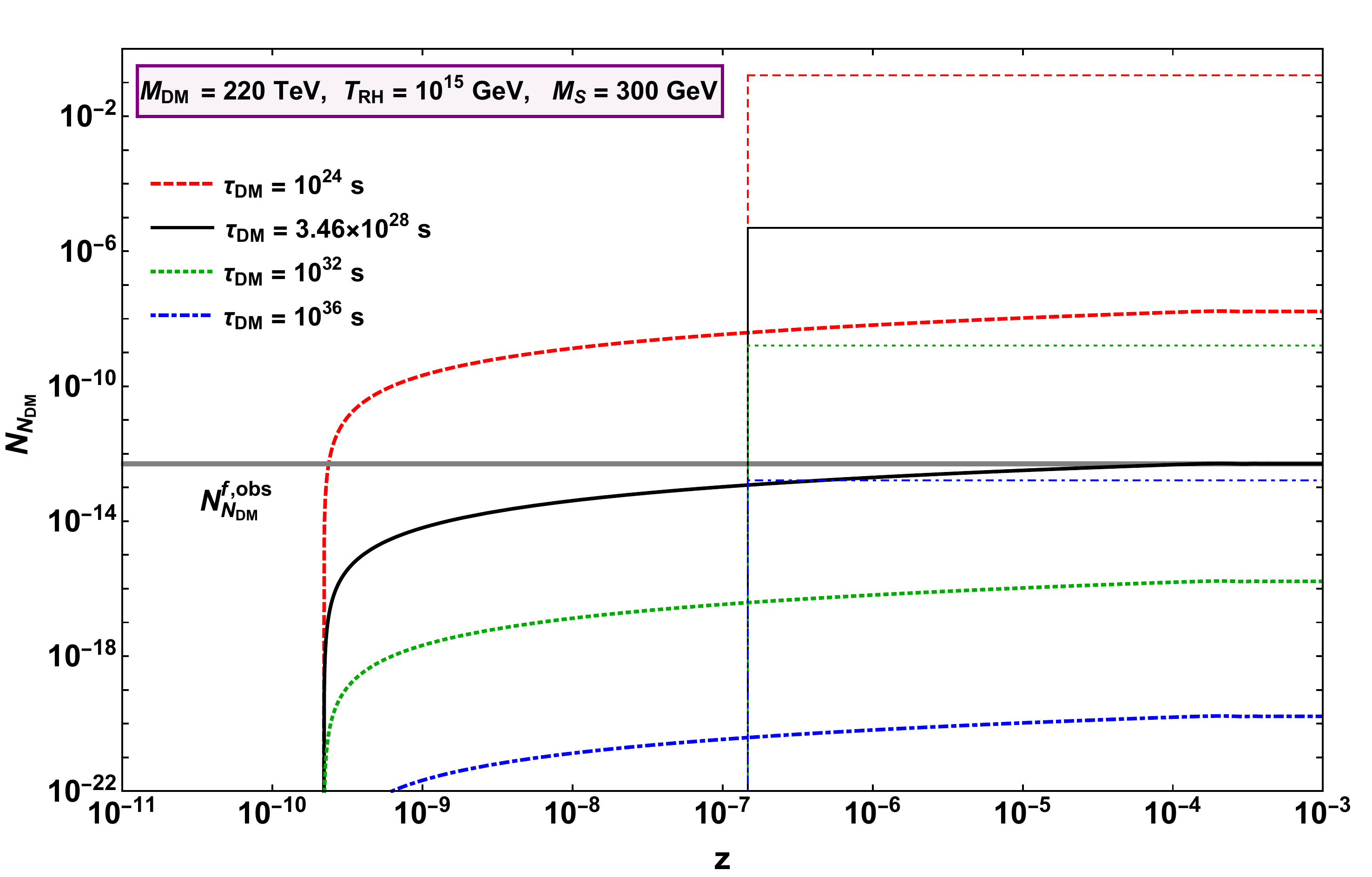,height=85mm,width=125mm} \\
\vspace{10mm}
\psfig{file=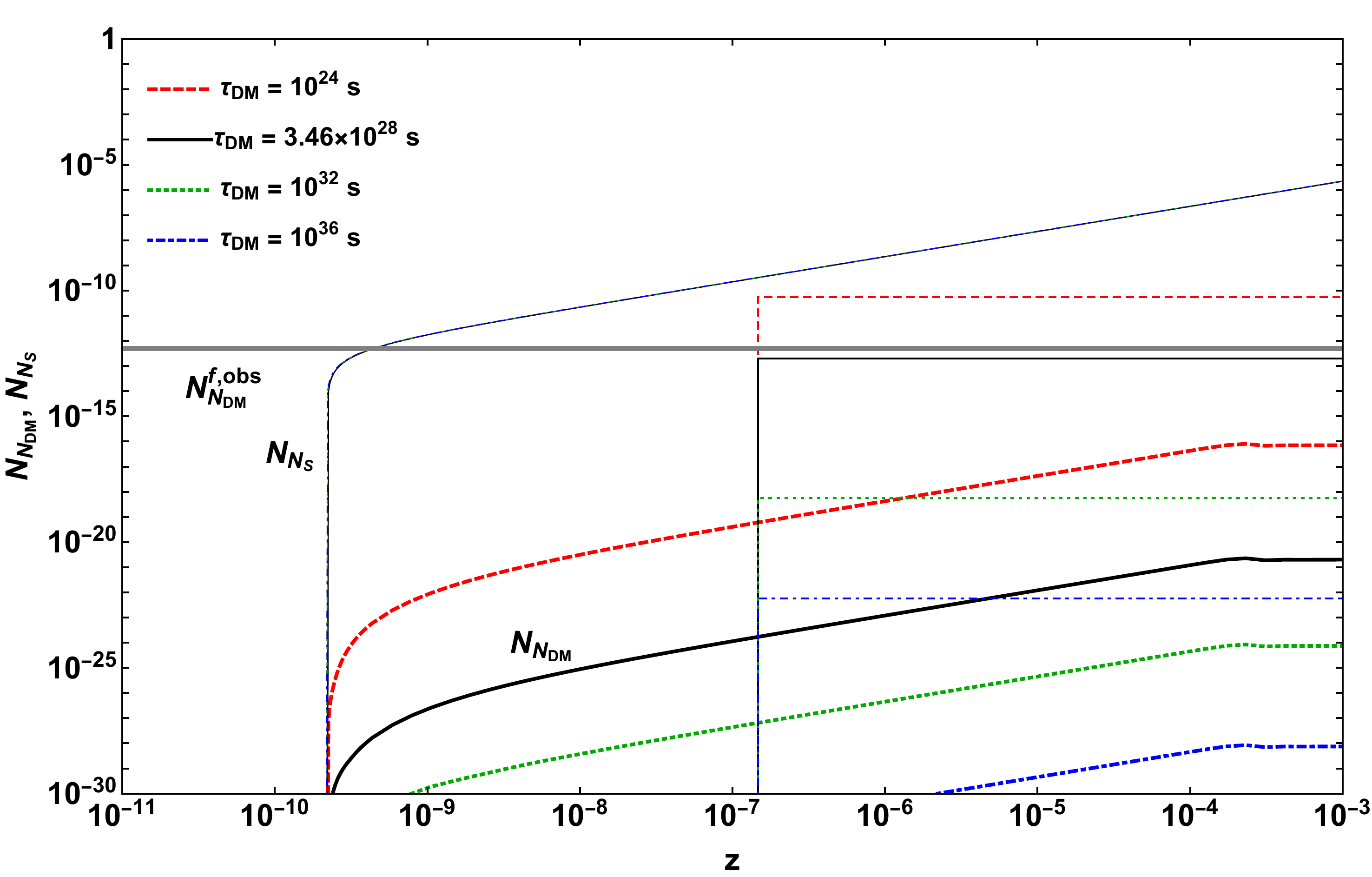,height=85mm,width=125mm}   
\end{center}
\vspace{-1mm}
\caption{Evolution of the DM abundance $N_{N_{\rm DM}}$ for different values
of $\tau_{\rm DM}$, as indicated, and for fixed values of $M_{\rm DM}, M_{\rm S}$ and $T_{\rm RH}$.
As in the previous figure, the upper (lower) panel is for initial thermal (vanishing) $N_{\rm S}$-abundance and
the step functions describe the instantaneous LZ approximation.}
\label{benchmark}
\end{figure}
It should be noticed that also in this case, for all values of $\tau_{\rm DM}$, 
the LZ approximation overestimates the abundance by about seven orders of magnitude.

\subsection{Dependence on $M_{\rm S}$}

Finally, in Fig.~4, we fix the value of $T_{\rm RH}=10^{15}\,{\rm GeV}$ 
and $\tau_{\rm DM}=3.46 \times 10^{28}\,{\rm s}$ while we show different 
evolutions of the $N_{\rm DM}$-abundance for different values of $M_{\rm DM}/M_{\rm S}$, or equivalently $M_{\rm S}$, explicitly: $M_{\rm DM}/M_{\rm S}=2.2 \times 10^5, 10^3, 10^2, 10, 1.1, 1.01$
corresponding respectively to $M_{\rm S}= 1 \,{\rm GeV}, 0.22\,{\rm TeV}, 2.2\,{\rm TeV}, 
22 \, {\rm TeV}, 200\,{\rm TeV}, 218 \, {\rm TeV}$.
Notice again how, 
in the case of initial vanishing $N_{\rm S}$-abundance, there is no value of $M_{\rm DM}/M_{\rm S}$
for which the relic $N_{\rm DM}$-abundance can reproduce the observed DM abundance, contrary
to the case of initial thermal $N_{\rm S}$-abundance. In the case of
initial vanishing $N_{\rm S}$-abundance and for $M_{\rm S} > M_W$, the relic $N_{\rm DM}$ abundance does not depend on 
$M_{\rm DM}/M_{\rm S}$ as far as  $M_{\rm DM}/M_{\rm S} \gg 1$, 
similarly to the independence of $T_{\rm RH}$ shown in Fig.~3. However, 
in the quasi-degenerate limit, for $M_{\rm DM} \simeq M_{\rm S}$, 
there is an increase of about one order of magnitude until full saturation.  
This is clearly in stark contrast with the LZ approximation, 
where increasing the value of $M_{\rm DM}/M_{\rm S}$ corresponds to an increased value of $T_{\rm res}$ 
and of the mixing angle. This translates into an increase of the relic abundance despite the fact that for higher temperature the value
of the $N_{\rm S}$-abundance at the resonance decreases. In the case of density matrix equation solutions, the freeze-in temperature
and the evolution $N_{\rm DM}(z)$ is approximately independent of $M_{\rm DM}/M_{\rm S}$ despite the fact that
the $N_{N_{\rm S}}$ is not, something that suggests that there is a compensation between higher mixing angle but smaller 
${N_{\rm S}}$-abundance for higher values of $M_{\rm DM}/M_{\rm S}$. Of course this compensation is absent assuming
initial thermal $N_{\rm S}$-abundance since this stays constant for $z\ll 1$ and in this way the relic abundance increases 
for increasing $M_{\rm DM}/M_{\rm S}$. 

Notice also how in the quasi-degenerate limit, for
$M_{\rm DM}/M_{\rm S} \ra 1$, the case as originally proposed in \cite{ad}, 
the result from the LZ approximation  tends toward the relic $N_{\rm DM}$ abundance
from the solution of density matrix equations though it is still two orders of magnitude higher,
both in the case of initial thermal and vanishing $N_{\rm S}$-abundance.
\begin{figure}
\begin{center}
\psfig{file=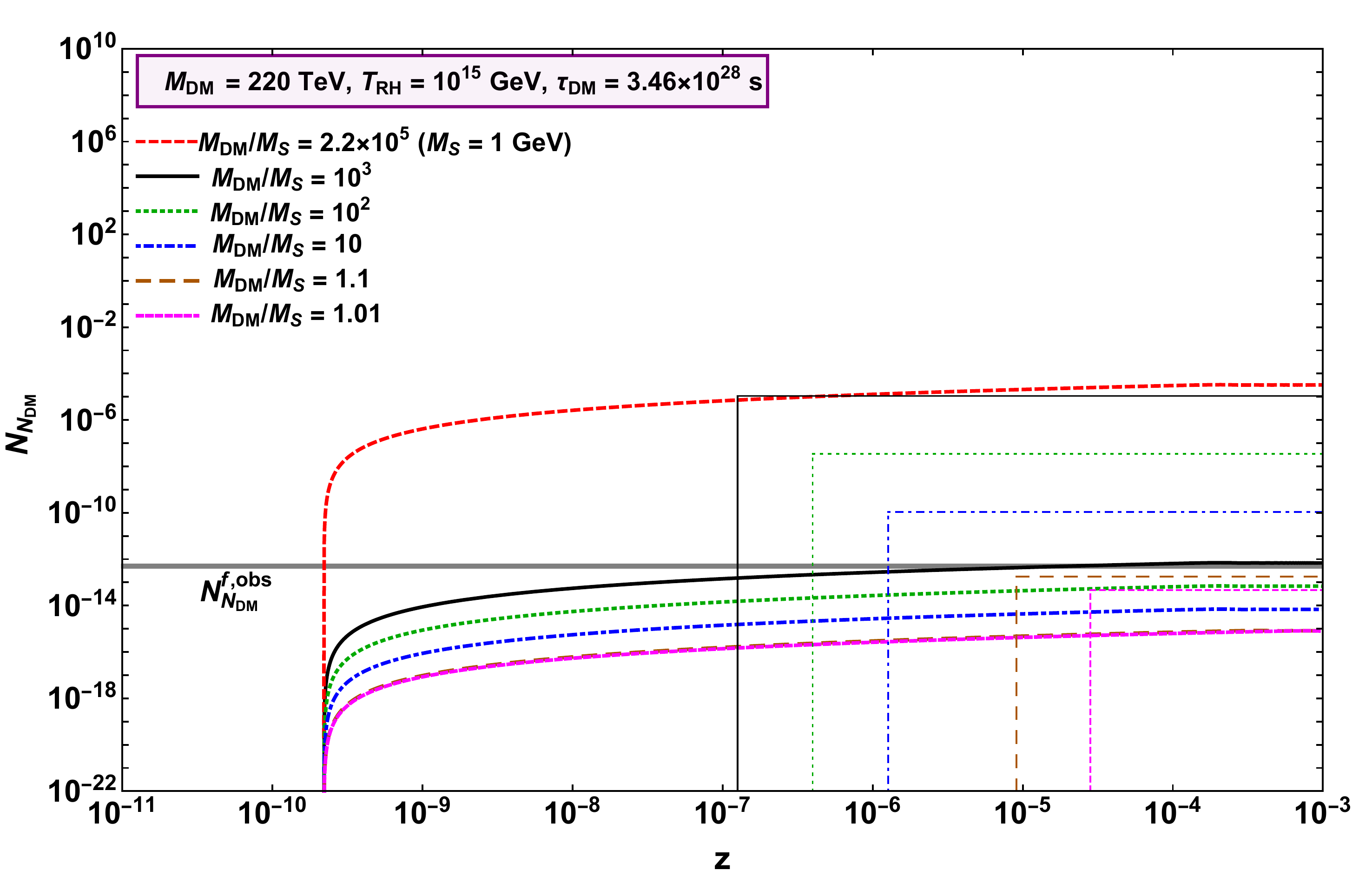,height=85mm,width=125mm} \\
\vspace{10mm}
\psfig{file=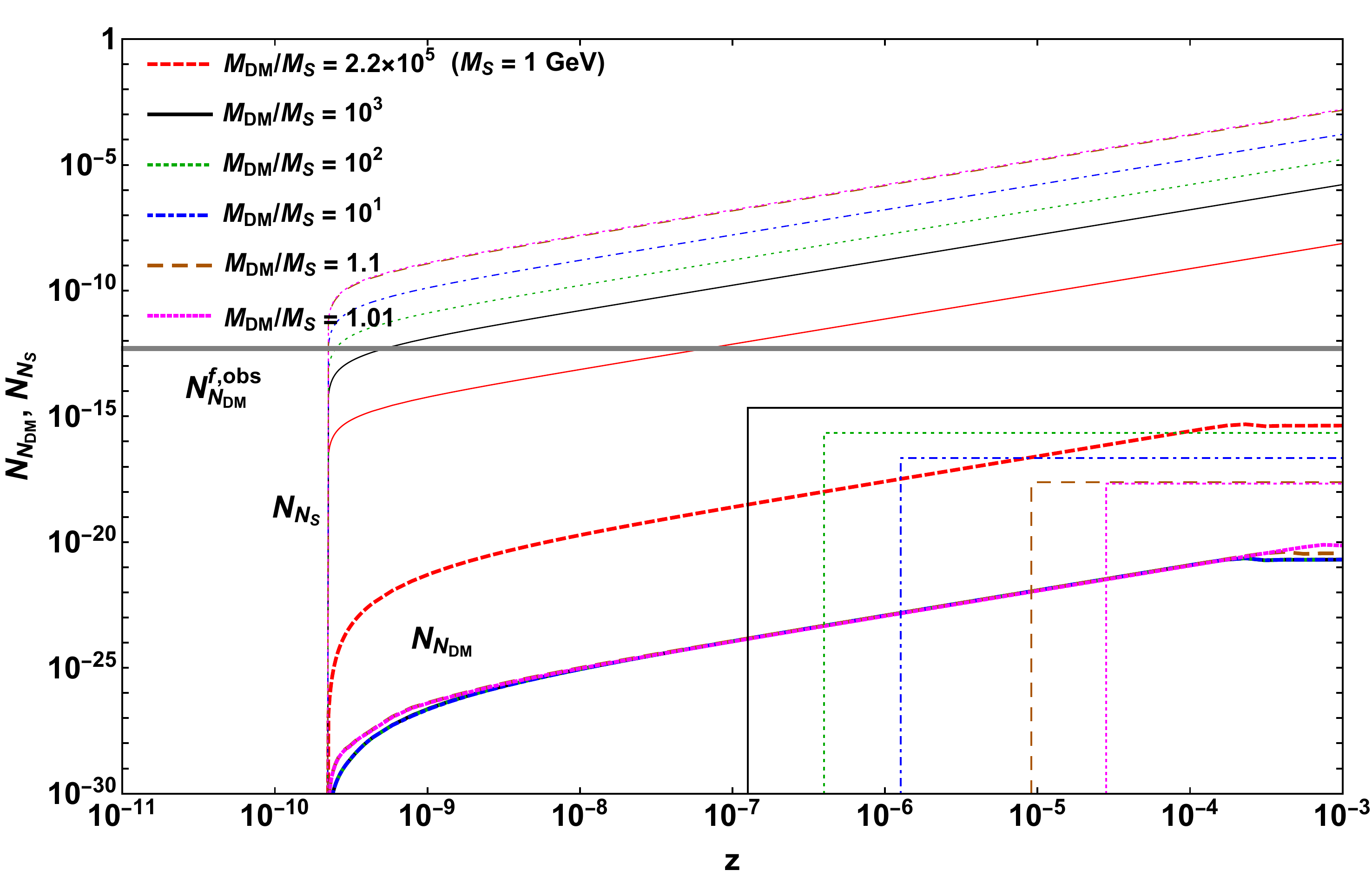,height=85mm,width=125mm}   
\end{center}
\vspace{-1mm}
\caption{Evolution of the DM abundance $N_{N_{\rm DM}}$ for different values
of $M_{\rm DM}/M_{\rm S}$, as indicated, and for fixed values of $T_{\rm RH}$ and $\tau_{\rm DM}$.
As in previous figures, the upper (lower) panel is for initial thermal (vanishing) $N_{\rm S}$ abundance and 
the step functions correspond to the the instantaneous LZ approximation.}
\label{benchmark}
\end{figure}
Finally, let us discuss the interesting case $M_{\rm S}=1\,{\rm GeV}$ (red lines).
In the hierarchical limit, for $M_{\rm DM}/M_{\rm S} \gg 1$, and in the LZ approximation 
the resonant temperature grows to very large values. In the case of initial vanishing
$N_{\rm S}$-abundance, the upper bound on the
reheat temperature translates into the upper bound on $M_{\rm DM}$  Eq.~(\ref{ubTRHvan}).
However, one can see that from the numerical solutions of the density matrix equation there is no resonant temperature and actually
most of the asymmetry is produced prior to  the freeze-in temperature that is independent
of $M_{\rm DM}/M_{\rm S}$. In this way the upper bound Eq.~(\ref{ubTRHvan}) does not
actually hold, and one can both lower $M_{\rm S}$ and increase $M_{\rm DM}$
in a way to suppress the four body decay rate for $M_{\rm S} < M_W$. 
When this happens, the same lifetime is obtained for a much lower value of $\widetilde{\L}$
(or equivalently higher value of the coupling $\la_{\rm DM-S}$) and this is why for $M_{\rm S} =1\,{\rm GeV}$
one can see that the relic abundance greatly increases.  For  $M_{\rm DM} =220\,{\rm TeV}$ this is still
not enough to reproduce the observed DM abundance in the case of initial vanishing $N_{\rm S}$-abundance.
However, as we will see, an allowed region at high values $M_{\rm DM} \gtrsim 20\,{\rm PeV}$, for $M_{\rm S} = 1\,{\rm GeV}$,
opens up also for vanishing initial  $N_{\rm S}$-abundance.

\subsection{Oscillations of the RH neutrino DM abundance prior to the freeze-in}

In Fig.~5, we show a log-linear plot of $N_{\rm DM}$ for initial thermal $N_{\rm S}$-abundance,
highlighting the oscillations of the DM abundance prior to the freeze-in.
In this case, we show the evolution $N_{N_{\rm DM}}(z)$ for fixed $M_{\rm S} =300\,{\rm GeV}$ 
but for three different choices of $T_{\rm RH}$ and $\tau_{\rm DM}$, 
in a way that the observed DM abundance is reproduced in all three cases. 
As one can see, we still choose the benchmark values $T_{\rm RH} =10^{15}\,{\rm GeV}$ and 
$\tau_{\rm DM} =3.46 \times 10^{28}\,{\rm s}$. 
\begin{figure}
\begin{center}
\psfig{file=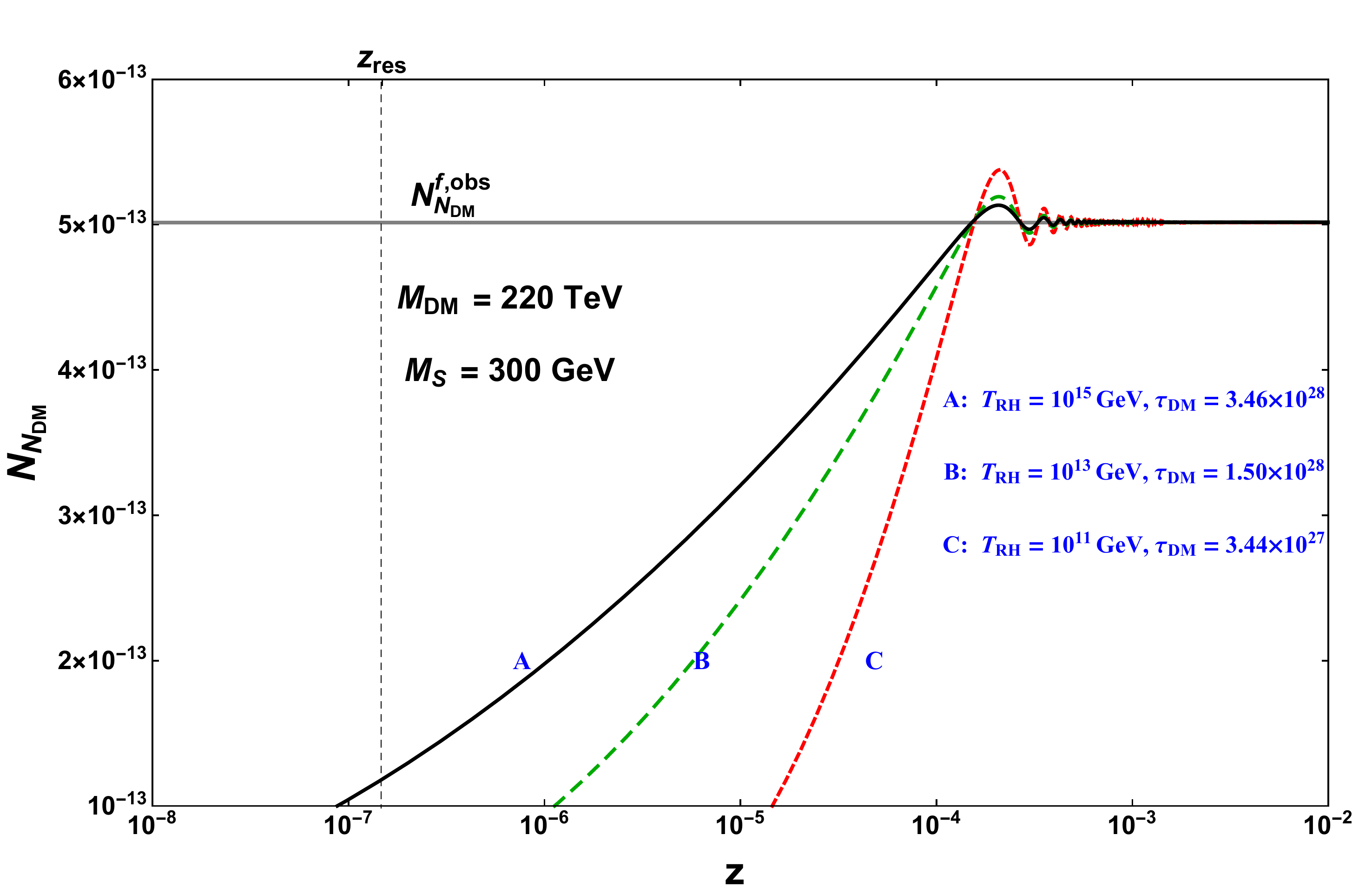,height=85mm,width=125mm} 
\end{center}
\vspace{-1mm}
\caption{Evolution of the DM abundance $N_{N_{\rm DM}}$ for three different choices
of $T_{\rm RH}$ and $\tau_{\rm DM}$ as indicated in a linear plot for the abundance.
For the three cases A, B and C one has respectively
$\widetilde{\L}=1.13\times 10^{24}\,{\rm GeV}, \widetilde{\L}=7.4 \times 10^{23}\,{\rm GeV}$ and 
$\widetilde{\L}=3.55 \times 10^{23}\,{\rm GeV}$.
}
\label{benchmark}
\end{figure}
The vertical line indicates the resonant temperature within the LZ approximation, 
and it can be noticed  again how this is much higher than the freeze-in temperature. 
For this reason, in the density matrix equation solution, though the production is much less efficient than
in the LZ approximation, this occurs at much lower temperatures and it allows to increase the scale of DM
and lower the scale of the source RH neutrino partly compensating the reduced efficiency. 

\subsection{Analytical insight}

Let us finally provide some analytical insight on the results we obtained, aiming especially at explaining why the LZ approximation
fails, overestimating by many orders of magnitude the relic dark matter abundance 
when $M_{\rm DM} \gg M_{\rm S}$.\footnote{Some of these considerations were anticipated in \cite{gws}.}
As one can clearly see  from Fig.~5, the problem with the LZ approximation is that it requires that  
neutrino oscillations have already developed at the resonance, while this is clearly not true. The DM production can then be 
explained in terms of neutrino oscillations occurring below the resonance, when the thermal mass from Yukawa interactions can be neglected in the diagonal terms in the Hamiltonian,  
but still with a time-dependent mixing angle.\footnote{If we again make an analogy
with left handed neutrino mixing and in particular with solar neutrino oscillations, below the resonance one
would recover vacuum neutrino oscillations. In our case thermal effects from the Anisimov operator are still
important in generating the time-dependent mixing angle and so it would not be correct
to say that one recovers `vacuum oscillations' but the analogy can be helpful to understand what happens.} An exhaustive description will
be given in a forthcoming paper \cite{preparation}. Here we just notice that  the 
LZ approximation can be used reliably only imposing the condition that 
at the resonance neutrino oscillations have already developed , implying $z_{\rm res}\gg z_{\rm osc}$, 
where $z_{\rm osc}$ is the value of $z$ corresponding to an age of the Universe $t_{\rm osc} = 12\,\pi T_{\rm osc}/\D M^2$,
where $t_{\rm osc}$ is defined as the time when the first dip occurs.
Using $t_{\rm osc} = 1/(2\,H(z_{\rm osc}))$, 
one finds $z_{\rm osc} \simeq 11\,M_{\rm DM}/(M_{\rm Pl}\,\D M^2)^{1/3}$. 
For the example of Fig. 5, one indeed correctly finds $z_{\rm osc} \simeq 3\times 10^{-4}$. 
Imposing $z_{\rm res} \gg z_{\rm osc}$, one finds for the validity of the LZ approximation a very stringent
constraint $M_{\rm DM} - M_{\rm S} \ll 10^{-12}\,{\rm GeV}$. This is consistent with the results of Fig.~4 where the LZ
limit is approached but never recovered, since for $M_{\rm DM} =220\,{\rm TeV}$ one would need $M_{\rm DM}/M_{\rm S}-1 \ll 10^{17}$.

This simple analytical insight allows some interesting considerations: 
\begin{itemize}
\item The DM production is well explained by simple two neutrino mixing below the resonance with a rapidly decreasing mixing angle in a way that  an abundance of source RH neutrino oscillates into a DM neutrino  abundance but this does not oscillate back. This simple
mechanism is quite a novel simple production mechanism.
\item As in the case of traditional non-resonant active-sterile neutrino oscillations in the early Universe, the monocromatic approximation we 
adopted is expected to work quite well and full momentum dependent description is expected to produce just a small correction.
\item An interesting point is that since the production is non-resonant anyway, the case $M_{\rm DM} \ll M_{\rm S}$
should work similarly but this time with even less constrained value of $\Delta M^2$ and, interestingly, traditional two RH neutrino seesaw
models and leptogenesis with hierarchical masses might potentially work in explaining neutrino masses and mixing and matter-antimatter asymmetry.
This scenario is very interesting and will be explored in a forthcoming paper \cite{preparation}. 
\end{itemize}

\subsection{Unifying dark matter and leptogenesis}

Within the Higgs induced RHiNo DM model,  the explanation of the DM abundance can be combined
with an explanation of the  matter-antimatter asymmetry within leptogenesis, obtaining a unified picture
of neutrino masses, dark matter and leptogenesis \cite{unified}.
In this case, the source RH neutrino should interfere with a third RH neutrino species
and they should be quasi-degenerate in order for the $C\!P$ asymmetry to be resonantly enhanced
and have successful leptogenesis much below the lower bound of $10^{10}\,{\rm GeV}$ holding 
in the hierarchical case. The observed baryon-to-photon ratio is given by \cite{planck18}
\be
\eta^{\rm obs}_{B0} = (6.12 \pm 0.04) \times 10^{-10}  \,  .
\ee
Using the same normalisation as for the RH neutrino abundances, the 
final $B-L$ asymmetry is related to the baryon-to-photon ratio predicted by
leptogenesis simply by $\eta_{B0}^{\rm lep} \simeq 0.01 \, N_{B-L}^{\rm f}$,
so that $N_{B-L}^{\rm f,obs} \simeq 6.1 \times 10^{-8}$ is the final $N_{B-L}$ value
needed to reproduce the observed value of $\eta_{B 0}$.

The evolution of the $B-L$ asymmetry with temperature can be calculated
as the sum of six contributions both on the two heavy neutrino flavours,
the source and the interfering RH neutrinos, and on the three charged lepton
flavours considering that the asymmetry will be generated in the three flavoured regime.
We can then write  \cite{pedestrians,beyond,predictions,dirac}
\be\label{NBmLasy}
N_{B-L}(z) = \sum_{\a = e,\mu,\tau} \, \left(N_{\D_\a}^{\rm (1)}(z) + 
N_{\D_\a}^{\rm (2)}(z)\right) \,  ,
\ee
where $N_{\D_\a}^{\rm (1)}(z)$ and $N_{\D_\a}^{\rm (2)}(z)$
are the abundances of the flavoured asymmetries $\D_{\a} \equiv B/3 - L_{\a}$
generated by the lightest and next-to-lightest RH neutrino 
(the source RH neutrino can be either one or the other).
The flavoured asymmetries can be calculated as ($I=1,2$) 
\be\label{flavasym}
N_{\D_\a}^{(I)}(z) = \ve_{I \a} \, \kappa_{I\a}(z, K_I, K_{1\a} + K_{2\a}) \,  ,
\ee
where $\ve_{I\a}$  and $K_{I\a}$ are respectively the $C\!P$ flavoured asymmetry and
the flavoured decay parameter associated to the RH neutrino $N_I$,
while $\kappa_{I\a}(z,K_{1\a} + K_{2\a})$ is the efficiency factor at 
temperature $T = M_{\rm S}/z_{\rm S}$ and an analytical solution
of the Boltzmann equation gives 
\bea\label{kappaI}
\kappa_{I\a}(z,K_I,K_{1\a} + K_{2\a})  & = &
\int_{z_{\rm S}^{\rm in}}^{z_{\rm S}}\,dz_{\rm S}' \,
(D_I+S_I)\,\left[N_{N_I}(z_{\rm S}') - N^{\rm eq}_{N_I}(z_{\rm S}')\right] \,  \\ \nonumber
& & \times \exp\left[-\int_{z_{\rm S}^{\rm in}}^{z_{\rm S}'}\,dz_{\rm S}'' \, 
W_{\a}(z_{\rm S}'') \right]  \,  ,
\eea
where remember that $z_{\rm S}=z\,M_{\rm DM}/M_{\rm S}$ and
 where $W_{\a}(z_{\rm S})$ is the wash-out term acting on flavour $\a$ including inverse decays and $\D L=1$ scatterings. 
 Notice that in $D_I$ and $S_I$ we are implying a dependence on the total decay parameter $K_I = \sum_\a \, K_{I\a}$.
 We refer the reader to Appendix C for more details on the calculation of the 
 asymmetry and expressions of flavoured $C\!P$ asymmetries and
 flavoured decay parameters. Here we just notice that the 
 asymmetry $N_{B-L}(z)$ depends on the low energy neutrino parameters, including the low
 energy phases, the degeneracy $\delta_{\rm lep} \equiv |M_1 - M_2|/M_1$ and one complex angle 
 in the orthogonal matrix that parameterises the Dirac neutrino mass matrix.
 
For a specific choice of these parameters that satisfy
successful leptogenesis and such that the asymmetry is dominantly produced by the decays of the
source RH neutrinos (see Appendix C for details), 
we plot in Fig.~6 the evolution of the $B-L$ asymmetry $N_{B-L}(z)$. We also plot $N_{N_{\rm DM}}(z)$ 
for a choice of values of the  parameters in the Higgs induced RHiNo DM scenario
that also reproduce the correct observed DM abundance (the same values as in case A in Fig~5).
\begin{figure}
\begin{center}
\psfig{file=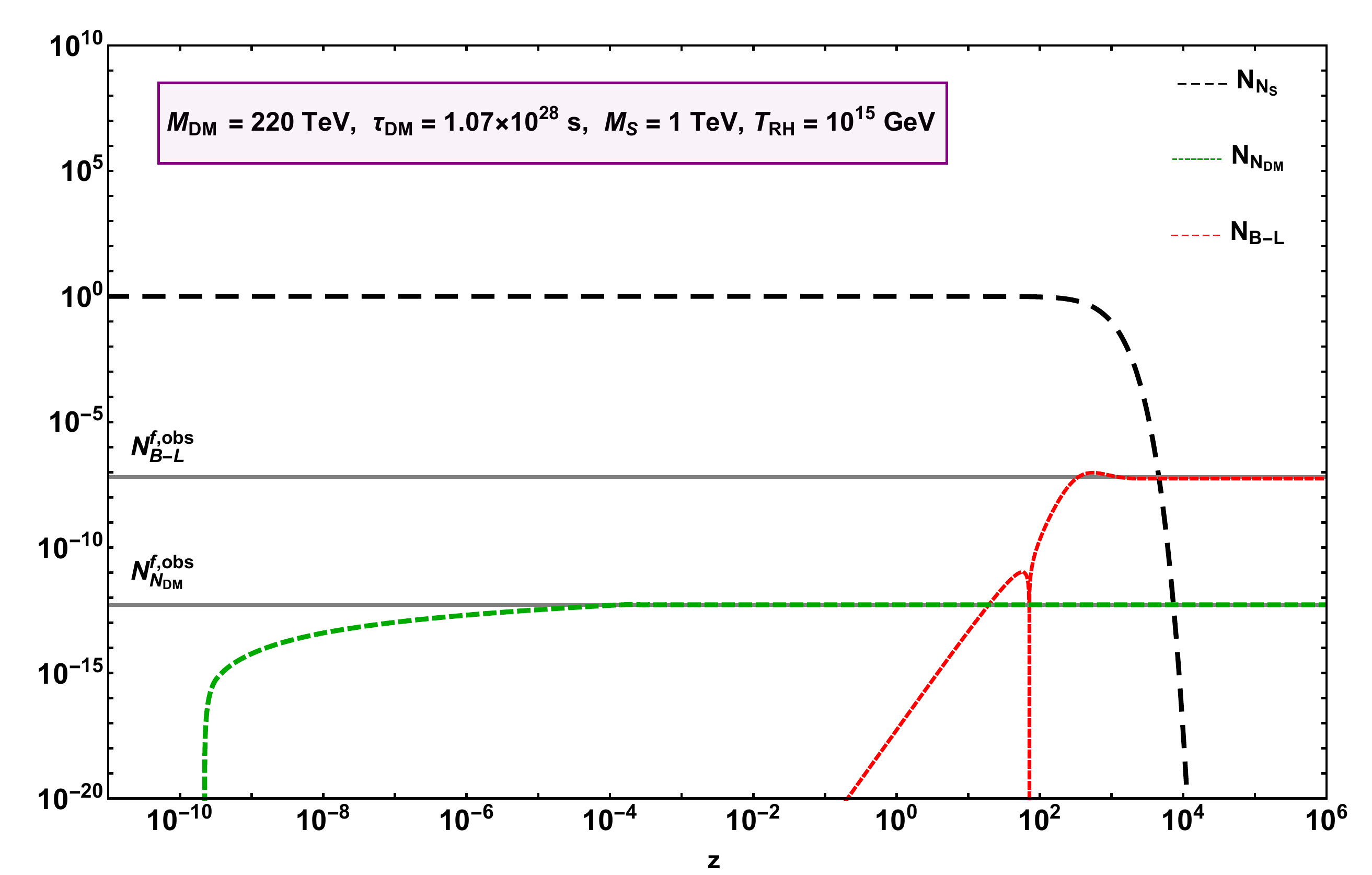,height=85mm,width=115mm} 
\end{center}
\vspace{-10mm}
\caption{Evolution of the $B-L$ asymmetry and 
DM abundance $N_{N_{\rm DM}}$ for a choice of parameters such that the final values simultaneously
reproduce the observed values of matter-antimatter asymmetry (see Appendix C) and DM abundance. 
The $N_{\rm S}$-abundance is also shown and one can see that the plot 
is for initial thermal $N_{\rm S}$-abundance.}
\label{benchmark}
\end{figure}

\section{Bounds on the DM mass}

In Fig.~7 we summarise the results we found for different choices of the parameters plotting
the allowed regions in the $M_{\rm DM}$-$\tau_{\rm DM}$ plane. In the higher panel we imposed 
the most conservative upper bound $T_{\rm RH} < 10^{15}\,{\rm GeV}$, in the central panel 
we imposed $T_{\rm RH} < 10^{12} \,{\rm GeV}$
and finally in the bottom panel we set more stringently $T_{\rm RH} < 10^{10}\,{\rm GeV}$.
\begin{figure}
\begin{center}
\psfig{file=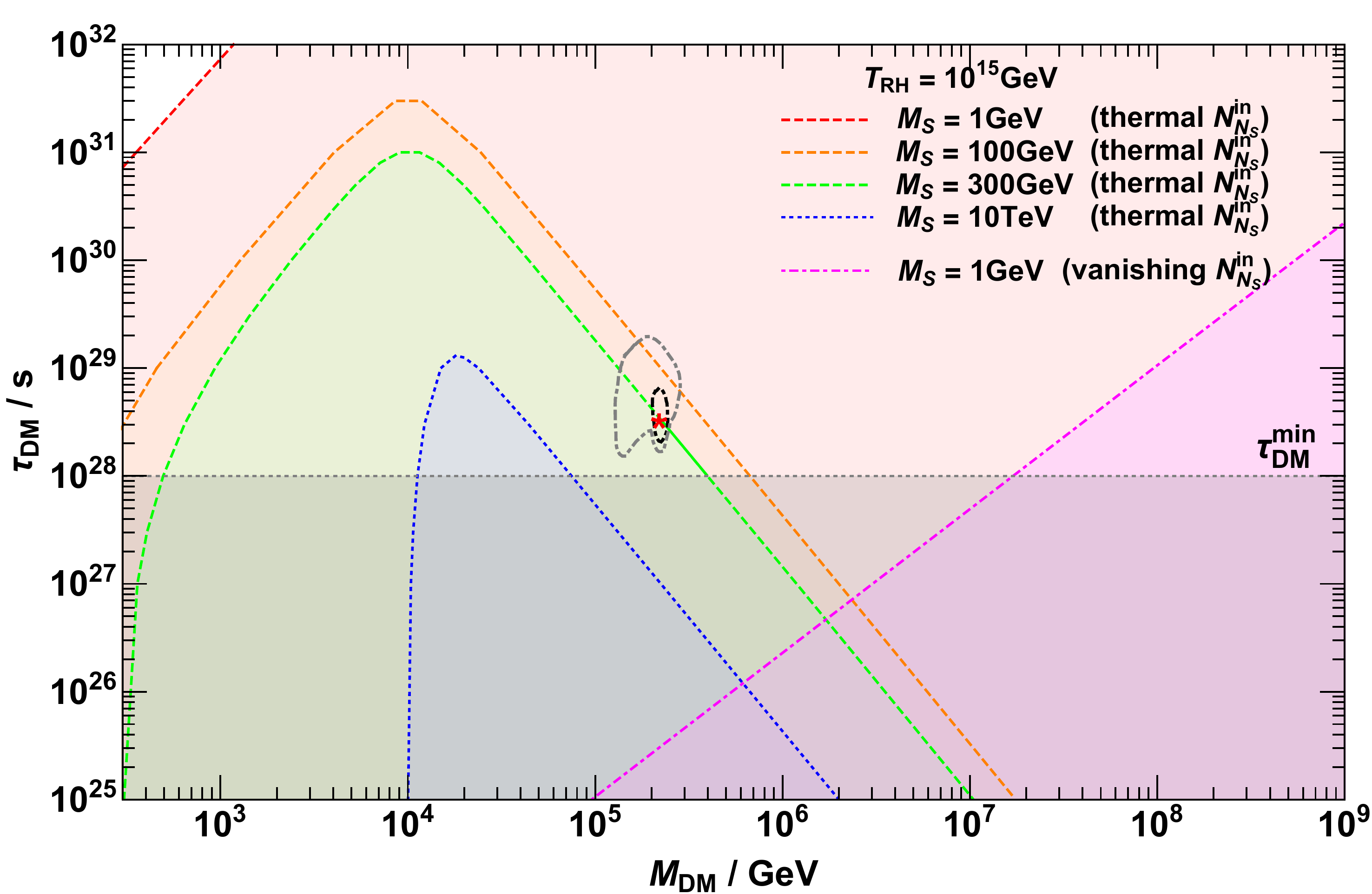,height=65mm,width=125mm} \\
\vspace{1mm}
\psfig{file=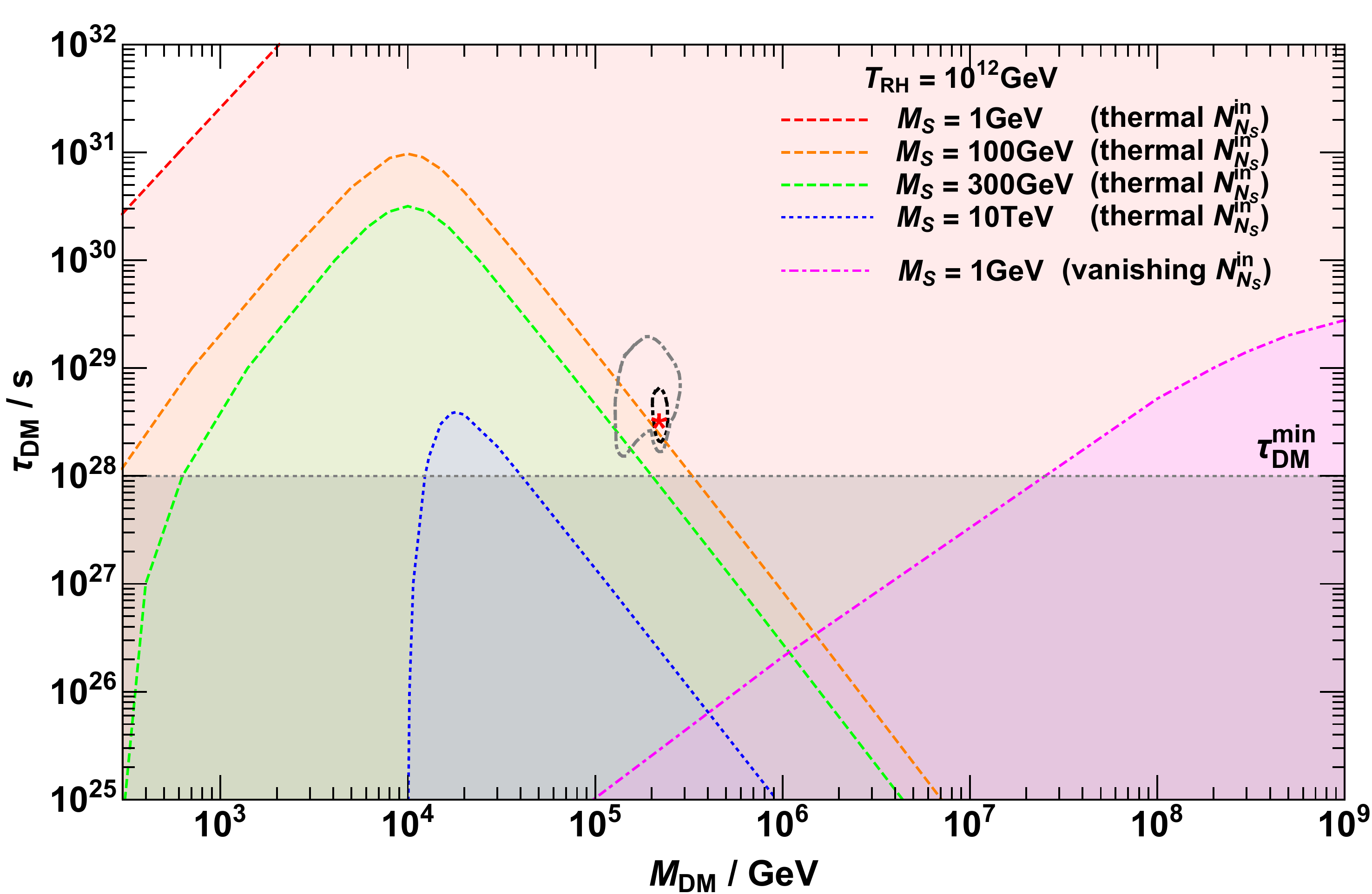,height=65mm,width=125mm} \\
\vspace{1mm}
\psfig{file=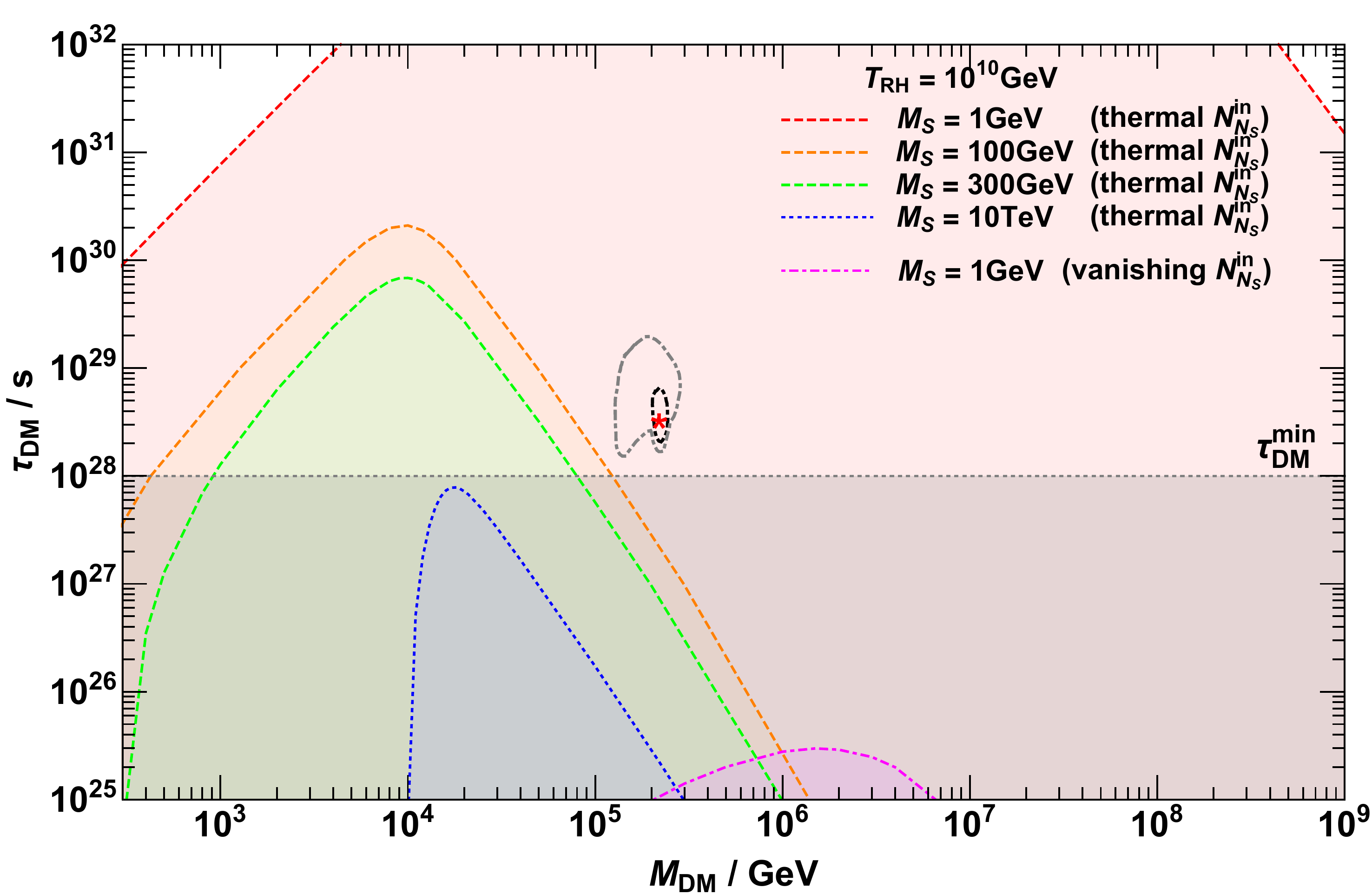,height=65mm,width=125mm}
\end{center}
\vspace{-10mm}
\caption{Allowed regions in the plane $\tau_{\rm DM}$ versus $M_{\rm DM}$ for different conditions as indicated.
The three panels correspond to three different choices of the upper bound on $T_{\rm RH}$. The red star and the contour lines around are
the same as in Fig.~1.}
\label{benchmark}
\end{figure}
In the upper panel, for $T_{\rm RH}< 10^{15}\,{\rm GeV}$, one can see how the
only way to have an allowed region for vanishing initial $N_{\rm S}$-abundance is
for $M_{\rm S} < M_W$ and in particular we show the allowed region for $M_{\rm S} > 1\,{\rm GeV}$.
As we discussed, this has the effect to suppress the four body decay rate, nullifying the upper bound on $M_{\rm DM}$.
However, one can see how in this case there is a lower bound $M_{\rm DM} \gtrsim 20\,{\rm PeV}$. 

On the other hand, for initial thermal $N_{\rm S}$-abundance, values for $M_{\rm S} > M_W$ and even values
$M_{\rm S}> 300\,{\rm GeV}$ compatible with a traditional scenario of leptogenesis from decays,  allowed regions
exist. In particular for $M_{\rm S} > 300\,{\rm GeV}$, one has 
$0.5\,{\rm TeV} \lesssim M_{\rm DM} \lesssim 0.5\,{\rm PeV}$ and lifetimes
as large as $10^{31}\,{\rm s}$ are allowed. One should appreciate how improvement in  the 
lower bound on $\tau_{\rm DM}$ from neutrino telescope experiments will progressively test 
the scenario placing more and more stringent constraints. 

In the central panel, for $T_{\rm RH} < 10^{12}\,{\rm GeV}$, there is no significant reduction of the allowed regions
and this is in line with what we noticed in Fig.~3: most of the DM abundance is produced prior to 
the freezing at $T_{\rm f} \sim 10^9 \,{\rm GeV}$ and therefore only when $T_{\rm RH}$ gets closer to $10^9\,{\rm GeV}$
one has a noticeable reduction of the relic DM abundance.

And indeed one can see that in the lower panel, for $T_{\rm RH} < 10^{10}\,{\rm GeV}$, all allowed regions shrink  considerably and, in particular, there is no allowed region for initial $N_{\rm S}$-abundance even for $M_{\rm S}> 1\,{\rm GeV}$. 
This stringent upper bound on the reheat temperature might be motivated for example by a supersymmetric version of the scenario, 
requiring an avoidance of the gravitino problem \cite{gravitino}.

\section{Final discussion}

We studied the production of the DM abundance within the Higgs induced RHiNo DM model,
solving numerically density matrix equation. The results show that the LZ approximation overestimates the 
DM abundance by many orders of magnitude. In the quasi-degenerate limit the mismatch is minimum but still the
DM abundance is overestimated by two orders of magnitude. It is then clear that a calculation employing density matrix equation
is crucial. Moreover the DM production occurs at temperatures much below the resonant temperature and this 
allows to open solutions for low values of the source RH neutrino mass.
We have seen that in this way solutions for initial vanishing $N_{\rm S}$-abundance are still possible but only for 
$M_{\rm S}$ below the $W$ boson mass and with a stringent lower bound on $M_{\rm DM}$. 
In particular, imposing $M_{\rm S}> 1\,{\rm GeV}$, we obtained $M_{\rm DM} \gtrsim 20\,{\rm PeV}$. 
In this case, one cannot reproduce the matter-antimatter asymmetry within traditional leptogenesis from decays, but it poses
the question whether this can be achieved considering leptogenesis 
from RH neutrino mixing that works indeed  for GeV RH neutrino masses \cite{ARS}. 

If one wants $M_{\rm DM} \sim 100\,{\rm TeV}$, as IceCube data seem to favour, 
these results then motivate the possibility to consider processes able to thermalise the source RH neutrino
prior to the DM abundance freezing. The existence of such processes is certainly plausible if one thinks that the non-renormalisable interactions in any case requires a UV-completion. One can for example think that the RH neutrinos at high temperatures might have extra-gauge interactions
and can get produced by very heavy $Z'$ bosons, a well known possibility \cite{plumacher}. However, even more interestingly, one can think
that the Higgs induced interactions for the source RH neutrinos are actually much stronger than for the DM RH neutrino and able to
thermalise the source RH neutrinos prior to the DM production. This possbility is quite attractive since it would not require additional
interactions. 

There is also another intriguing possibility emerging from our study. Within the LZ approximation it was necessary to
impose $M_{\rm DM} > M_{\rm S}$ in order to have a resonance. However, the numerical solution of the density matrix equations
show that the DM production is actually non-resonant. One can then wonder whether solutions with $M_{\rm DM} < M_{\rm S}$
might open up. These would be quite interesting since in this case the DM RH neutrino would be the lightest RH neutrino and one could embed
the mechanism within traditional two RH neutrino high energy scale seesaw models. 

Of course it would be also desirable to have an analytic understanding of our numerical results.\footnote{These have been cross-checked solving independent codes.} In particular, it would be
quite useful to have an analytic expression for  the final relic DM abundance and for the freeze-in temperature. 

Our results also should be generalised taking into account the momentum distribution. However, since RH neutrinos do not contribute to the effective potentials, complicated  back-reaction effects are excluded and including the momentum dependence should produce only corrections.  

In conclusion, a density matrix calculation of the Higgs induced RHiNo DM relic abundance is certainly necessary and confirms 
that the mechanism can reproduce  the observed DM abundance and simultaneously 
the matter-antimatter asymmetry within certain allowed regions in the space of parameters.  However, it also paves the way for new interesting scenarios, 
motivating further investigation in different directions. 

\vspace{-5mm}
\subsection*{Acknowledgments}

We wish to thank Teppei Katori for useful discussions.
PDB and YLZ acknowledge financial support from the STFC Consolidated Grant L000296/1. 
RS is supported  by a Newton International Fellowship (NF 171202) from Royal Society (UK) and SERB (India). 
KF acknowledges financial support from the NExT/SEPnet Institute.
This project has received funding/support from the European Union Horizon 2020 research and innovation 
programme under the Marie Sk\l{}odowska-Curie grant agreements number 690575 and  674896.

\newpage
\section*{Appendix A: Two body decay rate}

\appendix
\renewcommand{\thesection}{\Alph{section}}
\renewcommand{\thesubsection}{\Alph{section}.\arabic{subsection}}
\def\theequation{\Alph{section}.\arabic{equation}}
\renewcommand{\thetable}{\arabic{table}}
\renewcommand{\thefigure}{\arabic{figure}}
\setcounter{section}{1}
\setcounter{equation}{0}

In this Appendix we first derive the mixing angle induced by the Anisimov operator 
between the DM and the source RH neutrino at zero temperature Eq.~(\ref{thetaL0})
and responsible for the two body decay channel and then the resulting DM life time
when two body decays dominate.
After electroweak spontaneous symmetry breaking the Lagrangian
Eq.~(\ref{lagrangian}) becomes
\be
- {\cal L}^{\ell + \nu}_{m} = \overline{{\a}_{L}}\, m_\a \,{\a}_{R} +  \overline{\nu_{L \a }}\, m_{D \a J} \,  N_{R J} 
+{1\over 2}\,\overline{N_{R J}^{\,c}}\,M_J\, N_{R J} 
+ {1 \over 2} \, \overline{N_{R I}^c} \, \delta M^{\L}_{I J} \, N_{R J} + {\rm h.c.}  \;\;  ,
\ee
where $\delta M_{\L} =  {2\,\la_{IJ}\,v^2 /\L}$ is the effective Majorana mass term 
correction generated by  the Anisimov operator at zero temperature 
introducing off-diagonal terms in the total Majorana mass term 
\be
M^\L_{IJ} = M_{II}\,\d_{IJ} +  \delta M_{\L}  \,  .
\ee
This can be (Takagi) diagonalised by a unitary matrix $U_R^{\L\,T}$, in a way that 
\be\label{MIJL}
M_{IJ}^\L = U_R^{\L\,T}\, D_{M^\L} \, U_R^\L \,  ,
\ee
where $D_{M^\L} \equiv {\rm diag}(M_1',M_2')$, and $N_{R\,I} \rightarrow 
N'_{R J'} = U^\dagger_{R J' I}\, N_{R I}$.
Since we are in a two-neutrino mixing case this matrix can be made real 
and parameterised in terms of just one mixing angle, explicitly
\be\label{URL}
U_R^{\L}(\theta_\L^0) =
\left( \begin{array}{cc}
\cos\theta_{\L}^0 &  \sin\theta_{\L}^0 \\[1ex]
 -\sin\theta_{\L}^0   &  \cos\theta_{\L}^0 
\end{array}\right)  \,   .
\ee
The mixing angle can be easily calculated from (\ref{MIJL}).
The correction to the masses is negligible and one can approximate  $M'_I \simeq M_I$.
The important point is that in the new primed basis of mass eigenstates the 
neutrino Dirac mass matrix becomes $m'_{D \a I'} = m_{D \a J} \, U_{R J I'}$ 
and so using the parameterisation (\ref{URL}) one finds 
\be\label{URL}
 m'_{D \a I'} \simeq
\left( \begin{array}{cc}
-\theta_{\L}^0\,m_{De{\rm S}}    &  m_{De{\rm S}} \\[1ex]
  -\theta_{\L}^0\,m_{D\mu{\rm S}}  &   m_{D \mu{\rm S}}\\
  -\theta_{\L}^0\,m_{D\tau{\rm S}}  & m_{D \tau{\rm S}} 
\end{array}\right)  \,   ,
\ee
showing that the Higgs induced interactions generate small effective Yukawa couplings 
in the DM mass eigenstate that induce eventually its decays with a decay rate 
$\Gamma_{DM \ra A + \nu_{\rm S}} = \theta_{\L 0}^2\,h^2_{\rm S}\,M_{\rm DM}/(4\,\pi)$,
that of course vanishes in the limit $\widetilde{\L} \ra \infty$.

\newpage
\section*{Appendix B: Equivalent forms for the density matrix equation}

\renewcommand{\thesection}{\Alph{section}}
\renewcommand{\thesubsection}{\Alph{section}.\arabic{subsection}}
\def\theequation{\Alph{section}.\arabic{equation}}
\renewcommand{\thetable}{\arabic{table}}
\renewcommand{\thefigure}{\arabic{figure}}
\setcounter{section}{2}
\setcounter{equation}{0}

Let us show how from Eq.~(\ref{densitymatrixeq}) for the density matrix equation, 
one can easily arrive to the equivalent form Eq.~(\ref{densitymatrixeqpauli}).
First, using Eq.~(\ref{DH}), we can express the anti-commutator as 
\be
-i\,[{\cal H}, N]_{I J} = {P_0 \over 2} \, (\vec{V} \times \vec{P}) \cdot \vec{\sigma} \,  .
\ee
Defining then $R \equiv (\G_D+\G_S)\,(N_{N_{\rm S}}-N_{N_{\rm S}}^{\rm eq})$,
the second term in the right-hand side of Eq.~(\ref{densitymatrixeq}) can be recast as
\be
\begin{pmatrix}
0   &  {1\over 2}(\G_D+\G_S) \,N_{\rm DM-S}  \\ 
{1\over 2}(\G_D+\G_S)  \,N_{\rm S-DM}  & (\G_D+\G_S)\,(N_{N_{\rm S}}-N_{N_{\rm S}}^{\rm eq})  
\end{pmatrix}
={P_0\over 4}(\G_D+\G_S)\,\vec{P}_T\cdot \vec{\sigma} + 
\begin{pmatrix}
0   &  0  \\ 
0 & R
\end{pmatrix}
\ee
and also one can write
\be
\begin{pmatrix}
0   &  0  \\ 
0 & R
\end{pmatrix} = 
{R \over 2}\, I - {R \over 2}\,\sigma_z  \, .
\ee
With some straightforward steps one then arrive to 
\be
{d \vec{P} \over dt} \cdot \vec{\sigma} = (\vec{V} \times \vec{P}) \cdot \vec{\sigma} 
- {1\over 2}\,(\G_D+\G_S)\, \vec{P}_T\cdot \vec{\sigma} + {1\over P_0} \, \left( R - {dP_0 \over dt}\right)
- {1\over P_0}\,\left(R\,\sigma_z + {dP_0 \over dt}\, \vec{P}\cdot \vec{\sigma} \right) \,  ,
\ee
implying $R = dP_0 /dt$ and Eq.~(\ref{densitymatrixeqpauli}).

Finally, it is easy to see that Eq.~(\ref{densitymatrixeq}) is fully equivalent to another
popular form for the density matrix equation expressed in terms of anti-commutators \cite{siglraffelt,ARS}
\be
{dN_{IJ} \over dt} = -i\,[{\cal H}, N]_{IJ}  
- {1\over 2}\,\left\{\Gamma, N \right\}_{IJ} + {1\over 2}\, \left\{\Gamma, N^{\rm eq} \right\}_{IJ} \,  ,
\ee
where simply
\be
\G=\begin{pmatrix}
0   &  0  \\ 
0 & \G_D + \G_S
\end{pmatrix} 
\hspace{5mm} \mbox{\rm and} \hspace{5mm}
N^{\rm eq}=\begin{pmatrix}
N_{\rm DM}^{\rm eq}   &  0  \\ 
0 & N_{\rm S}^{\rm eq}
\end{pmatrix}    \;   .
\ee

\newpage
\section*{Appendix C: leptogenesis with two quasi-degenerate RH neutrinos}

\renewcommand{\thesection}{\Alph{section}}
\renewcommand{\thesubsection}{\Alph{section}.\arabic{subsection}}
\def\theequation{\Alph{section}.\arabic{equation}}
\renewcommand{\thetable}{\arabic{table}}
\renewcommand{\thefigure}{\arabic{figure}}
\setcounter{section}{3}
\setcounter{equation}{0}

In this Appendix we give more details on the calculation of the $B-L$ asymmetry within a two quasi-degenerate RH neutrino
scenario and  the specific case  shown in Fig.~6. The flavoured decay parameters are defined as 
\be
K_{I\a}\equiv \frac{\G_{I\a} + \overline{\G}_{I\alpha}}{H(T=M_I)} = \frac{|m_{D\a I}|^2}{M_I \, m_{\star}}  \,   ,
\ee
where the equilibrium neutrino mass 
\be
m_{\star} \equiv \frac{16\,\pi^{5/2}\,\sqrt{g_{\star}^{\rm SM}}}{3\,\sqrt{5}} \, 
\frac{v^2}{M_{\rm Pl}} \simeq 1.1 \,\, {\rm meV} \,  .
\ee
The flavoured $C\!P$ asymmetries are defined as
\be
\ve_{I\a}\equiv - \frac{\G_{I\alpha}-\overline{\G}_{I\alpha}}{\G_{I}+\overline{\G}_{I}} 
\simeq \frac{\overline{\ve}(M_I)}{K_I}
\, \left\{ {{\cal I}_{IJ}^{\a}}\,\x({M^2_J/ M^2_I}) + 
{{\cal J}_{IJ}^{\a}} \, \frac{2}{3(1-M^2_I/M^2_J)}\right\} \,  ,
\ee
where
\be
\overline{\ve}(M_I) \equiv \frac{3}{16\,\pi} \, \left(\frac{M_I\,m_{\rm atm}}{v^2}\right) 
\simeq 1.0 \times 10^{-13} \, \left(\frac{M_I}{{\rm TeV}}\right)  \,  ,
\ee
while ${\cal I}_{IJ}^{\a}$ and ${\cal J}^\a_{IJ}$ originate from the 
interference of tree level with one loop self-energy and vertex-diagrams and are given by
\be
{\cal I}_{IJ}^{\a} \equiv  \frac{{\rm Im}\left[m_{D\a I}^{\star} \, m_{D\a J} \, (m_D^{\dag} \, m_D)_{IJ}\right]}{M_I \, M_J \, m_{\rm atm} \, m_{\star} } ~,
\hspace{5mm}
{\cal J}_{IJ}^{\a} \equiv  
\frac{{\rm Im}\left[m_{D\a I}^{\star} \, m_{D\a J} \, (m_D^{\dag} \, m_D)_{JI}\right]}{M_I\, M_J \, m_{\rm atm} \, m_{\star} } \, \frac{M_I}{M_J}    \,   .
\ee
In the case of two RH neutrinos with quasi-degenerate masses, Eq.~(\ref{NBmLasy}) gives for the
final $B-L$ asymmetry \cite{unified}
\be\label{NBmLfin}
N_{B-L}^{\rm f} \simeq \frac{\overline{\ve}(M_1)}{3\,\d_{\rm lep}}
\left(\frac{1}{K_1} + \frac{1}{K_2}\right)\,
\sum_{\a}\,\k^{\rm f}(K_{1\a}+K_{2\a})\,\left[{\cal I}_{12}^{\a}+ {\cal J}_{12}^{\a}\right]  \,  ,
\ee
where $\k^{\rm f}$ is the final value of the efficiency factor that can be obtained from Eq.~(\ref{kappaI})
in the limit $z_{\rm S} \ra \infty$ (see discussion below for an analytical expression).
In order to take into account the low energy neutrino experimental data, it is convenient to
introduce the orthogonal parameterisation for the neutrino Dirac mass matrix \cite{ci},
\be
m_D=U\,\sqrt{D_m}\,\O\,\sqrt{D_M} \,   .
\ee
Since in our case $m_1 =0$, one has $m_2 = m_{\rm sol} \simeq 8.6 \,{\rm meV}$
and $m_3 = m_{\rm atm} \simeq 50\,{\rm meV}$, where $m_{\rm sol}$ 
and  $m_{\rm atm}$ are the solar and atmospheric neutrino mass scales
respectively and we consider only normal hierarchy since this is now quite strongly favoured by the 
global analyses compared to inverted hierarchy \cite{nufit}.

The leptonic mixing matrix can be parameterised in terms of three mixing angles and three $C\!P$ violating phases as
\begin{equation}
U = \left(
\begin{array}{ccc}
c_{13} c_{12} &
c_{13} s_{12} &
s_{13} e^{-i \delta} \\
- c_{23} s_{12} - s_{23} s_{13} c_{12} e^{i \delta} &
c_{23} c_{12} - s_{23} s_{13} s_{12} e^{i \delta} &
s_{23} c_{13} \\
s_{23} s_{12} - c_{23} s_{13} c_{12} e^{i \delta} &
-s_{23} c_{12} - c_{23} s_{13} s_{12} e^{i \delta} &
c_{23} c_{13}
\end{array}
\right) \, \mathrm{diag}(e^{i\rho},\, e^{i\sigma},\, 1) ~,
\end{equation}
where $c_{ij} = \cos \theta_{ij}$, $s_{ij} = \mathrm{sin}\theta_{ij}$, $\delta$ is the Dirac phase 
and $\rho$ and $\sigma$ are the two Majorana phases. 
%
%
The best fit values and the $1\s$ ($3\s$) confidence level (C.L.) ranges of the reactor, solar, atmospheric mixing angles
and Dirac phase for normal hierarchy are given by~\cite{nufit}
\bea\label{eq:expranges}
\nonumber
\theta_{13} & = &  {8.61^{\circ}}^{+0.13^{\circ}}_{-0.13^{\circ}}  \;
(8.22^{\circ}\mbox{--}8.99^{\circ})   \,  ,
 \\ \nonumber
\theta_{12} & = &  {33.82^{\circ}}^{+0.78^{\circ}}_{-0.76^{\circ}} \;  
(31.61^{\circ}\mbox{--}36.27^{\circ})  \, ,
\\
\theta_{23} & = &  {48.3^{\circ}}^{+1.1^{\circ}}_{-1.9^{\circ}} \;   
 (40.8^{\circ}\mbox{--}51.3^{\circ})     \,  ,
\\
\d & = &  {-138^{\circ}}^{+38^{\circ}}_{-28^{\circ}} \; 
 (-159^{\circ}\mbox{--}+10^{\circ})  \,   ,
\eea 
while there is no experimental information on the Majorana phases. 
Coming finally to the orthogonal matrix, in the two RH neutrino case this is given by 
\be\label{orthogonal}
\O = \left(
\begin{array}{ccc}
   0  &  0    & 1  \\
    \cos \o  &   \sin \o     & 0  \\
 -\z \sin \o & \z \cos \o & 0 
\end{array}
\right)   \,  ,
\ee
where $\o$ is a complex angle and $\z = \pm 1$ is a discrete parameter and the two possible values correspond to two different distinct branches of $\Omega$, with positive and negative determinant respectively~\cite{2RHN}.

In terms of the orthogonal parameterisation, the flavour decay parameters, in
the considered two quasi-degenerate RH neutrino case, can be written as
\be
K_{I\a} = \left|\sum_j\,\sqrt{\frac{m_j}{m_{\star}}} \, U_{\a j} \, \O_{j I}\right|^2  = 
\left|\sqrt{K_{\rm sol}}\,U_{\a 2}\,\O_{2 I} +  \sqrt{K_{\rm atm}}\,U_{\a 3}\,\O_{3 I}\right|^2 \,  ,
\ee
where we defined $K_{\rm sol} \equiv m_{\rm sol}/m_{\star} \simeq 8$ and
$K_{\rm atm} \simeq m_{\rm atm}/m_{\star} \simeq 45$. Using Eq.~(\ref{orthogonal}) we can then
write them in terms of the complex angle, obtaining
\bea
K_{1\a} & = & \left|\sqrt{K_{\rm sol}}\,U_{\a 2}\,\cos\o -\zeta \,  \sqrt{K_{\rm atm}}\,U_{\a 3}\,\sin\o \right|^2 \,  , \\
K_{2\a} & = & \left|\sqrt{K_{\rm sol}}\,U_{\a 2}\,\sin\o +\zeta \,  \sqrt{K_{\rm atm}}\,U_{\a 3}\,\cos\o \right|^2  \, .
\eea
We can also re-write ${\cal I}_{12}^{\a}$ and ${\cal J}^\a_{12}$ in terms of the orthogonal parametrisation, obtaining
\bea
{\cal I}_{12}^{\a} & = &   \sum_{k,l,m =2,3} \, \frac{m_k\,\sqrt{m_l\, m_m}}{m_{\rm atm}\,m_{\star} }\,
\mathrm{Im}\left[
U^{\star}_{\a m}\,U_{\a l} \, \O^{\star}_{m 1}\,\O_{l 2}\, \O^{\star}_{k 1}\,\O_{k 2}\right]  ~, \\
{\cal J}_{12}^{\a} & = & \, \sum_{k,l,m=2,3} \, \frac{m_k\,\sqrt{m_l\, m_m}}{m_{\rm atm}\,m_{\star}}\,
\mathrm{Im}\left[
U^{\star}_{\a m}\,U_{\a l}\,
\O^{\star}_{m 1}\,\O_{l 2}\, \O^{\star}_{k 2}\,\O_{k 1}\right] ~. 
\eea
Therefore, one can see that the final asymmetry will depend on two light neutrino mass scales, five parameters in $U$ (1 Majorana
phase can be removed in the two RH neutrino case),  on the
common heavy neutrino mass scale $M_{\rm S}$, on the relative mass splitting $\d_{\rm lep}$ and
on the complex angle $\o$ (in total eleven parameters).
In order to obtain a specific solution, that we show in Fig.~6, we have taken a real $U$, 
and choose ${\rm Re}[\o]=0$ and ${\rm Im}[\o] \ll 1$.\footnote{This is actually a limit that is well motivated 
within models with discrete flavour symmetries: when the symmetry is conserved the orthogonal
matrix coincides exactly with the permutation matrix and there is no $C\!P$ violation at high energies,
when the symmetry is broken, the small breaking parameter can be identified with $|\o|$ and is responsible
for $C\!P$ violation \cite{jenkins,feruglio}.}
In this case
one obtains 
\be
{\cal I}_{12}^{\a} \simeq {\cal J}_{12}^{\a} \simeq {\rm Im}[\o] \, U_{\a 2}\,U_{\a 3}\, K_{\rm sol} \, 
\left(\sqrt{K_{\rm atm} \over K_{\rm sol}} + \sqrt{K_{\rm sol} \over K_{\rm atm}} \right)  \,  .
\ee
and
\be
K_{1\a}  \simeq  K_{\rm sol}\, U_{\a 2}^2 \, , \;\;\;
K_{2\a} \simeq   K_{\rm atm}\, U_{\a 3}^2   \,   ,
\ee
implying $K_1 \simeq K_{\rm sol}$ and $K_2 \simeq K_{\rm atm}$.
In this way the expression for the final asymmetry Eq.~(\ref{NBmLfin}) gets specialised into
\be\label{NBmLfin2}
N_{B-L}^{\rm f} \simeq {2\over 3} \,  \frac{{\rm Im}[\o]\, \overline{\ve}(M_{\rm S})}{\d_{\rm lep}}
\left(1 + \frac{K_{\rm sol}}{K_{\rm atm}}\right)\, 
\left(\sqrt{K_{\rm atm} \over K_{\rm sol}} + \sqrt{K_{\rm sol} \over K_{\rm atm}} \right) \,
\sum_{\a}\,\k^{\rm f}(K_{1\a}+K_{2\a})\, U_{\a 2}\,U_{\a 3}\,   .
\ee
In Fig.~6 we have taken the best fit values for the three mixing angles $M_{\rm S} = 1\,{\rm TeV}$,
${\rm Im}[\o] = 0.01$ and $\d_{\rm lep} \simeq 4 \times 10^{-10}$, 
the correct values  to reproduce $N_{B-L}^{\rm f,obs} \simeq 6.1 \times 10^{-8}$.
Notice that of course one could also have chosen a real $\o$ and in that case one would have obtained a solution
where all asymmetry stems from low energy phases, but in that case one needs ${\rm Re}[\o] \gtrsim 1$. 
However, also in the case of a purely imaginary $\O$ that we are considering, 
despite one has a vanishing total $C\!P$ asymmetry,  the final asymmetry does not, since
the different washout in the different flavours prevents a full cancellation of the flavoured asymmetries. 

Finally, let us briefly discuss the calculation of the efficiency factor at temperature $T=z_{\rm S}/M_{\rm S}$, 
introduced in Eq.~(\ref{kappaI}), and its final value $\kappa^{\rm f}$.\footnote{This discussion 
extends the results found in \cite{pedestrians,beyond,predictions,dirac} to the case of two quasi-degenerate RH neutrinos
including flavour effects.}

The efficiency factor is related to the $B-L$ asymmetry
simply by Eq.~(\ref{flavasym}). The expression (\ref{kappaI}) comes from the solution 
of the set of rate equations  ($I=1, 2; \a =e,\mu, \tau$)
\bea\label{flavors} \label{dNNI}
{dN_{N_I}\over dz_{\rm S}} & = & -(D_I+S_I)\,(N_{N_I}-N_{N_I}^{\rm eq}) \\  \label{dNDal}
{dN^{(I)}_{\D_{\a}}\over dz_{\rm S}} & = &
\ve_{I\a}\,(D_I+S_I)\,(N_{N_I}-N_{N_I}^{\rm eq}) -\,W_{\a}\,N_{\D_{\a}}^{(I)} \,   ,
\eea
where the source RH neutrino is either the lightest (${\rm S}=1$) or the next-to-lightest (${\rm S}=2$) one
and where let us recall that $z_{\rm S} \equiv M_{\rm S}/T = z \, M_{\rm S}/M_{\rm DM}$. Notice
that since the two RH neutrinos are quasi-degenerate, then $z_{\rm S} \simeq M_1 /T \simeq M_2/T$.
The equilibrium RH neutrino abundance $N_{N_I}^{\rm eq}$ is given by Eq.~(\ref{NNeq}) and it is
actually the same for lightest ($I=1$) and next-to-lightest ($I=2$) RH neutrino since they are quasi-degenerate.
The sum of the decay and $\D L =1$ scattering terms can be written as 
\be
D_I(z_{\rm S}) +S_I(z_{\rm S}) = D_I(z_{\rm s})\,j(z_{\rm S})  \,  ,
\ee
where 
\be
D_I(z_{\rm S}) = K_I \, z^2_{\rm S} \, {\int_{z_{\rm S}}^{\infty}\,dx \, \sqrt{x^2 - z_{\rm S}^2}\,e^{-x} \over 
\int_{z_{\rm S}}^{\infty}\,dx \, x \,  \sqrt{x^2 - z_{\rm S}^2} \, e^{-x}} 
\ee
accounts for decays 
while the function $j(z_{\rm S})$ accounts for $\D L =1$ scatterings and can be written as
\be
j(z_{\rm S}) \simeq \left[{0.2\over z_{\rm S}} + {z_{\rm S} \over a } \ln\left(1+{a \over z_{\rm S}}\right)\right]\,
\left(1 + {2 \over z_{\rm S}} \right) \,   ,
\ee
where $a \simeq 8\pi^2/[9 \,\ln(M_{\rm S}/M_H)]$ and $M_H \simeq 0.4 \, T$ is the thermal Higgs mass.
Notice that $j(z_{\rm S} \gg 1) \simeq 1$.
The wash-out terms  $W_{I\a}(z_{\rm S})$,  the sum of the inverse decays and $\D L=1$ scattering wash-out terms, 
can be also written through the function $j(z_{\rm S})$ as
\be
W_{\a}(z_{\rm S}) = j(z_{\rm S})\, W_{\a}^{ID}(z_{\rm S})  \,  ,
\ee
where the flavoured inverse decay wash-out rate can be expressed as
\be
W_{\a}^{ID}(z_{\rm S}) = {1 \over 2} \, (K_{1\a}+K_{2\a})\, z_{\rm S}^2 \, 
\int_{z_{\rm S}}^{\infty}\,dx \sqrt{x^2 - z_{\rm S}^2} \, e^{-x}\,  .
\ee
The integral expression Eq.~(\ref{kappaI}) for the efficiency factor is simply the solution, in the form
of a Laplace integral, of the rate equations (\ref{dNDal}) describing the evolution of the
flavoured asymmetries. Using the rate equation for the RH neutrino abundance Eq.~(\ref{dNNI}),
Eq.~(\ref{kappaI}) for the efficiency factor can be recast as
\be\label{kappaIbis}
\kappa_{I\a}(z_{\rm S},K_I,K_{1\a} + K_{2\a})   = 
-\int_{z_{\rm S}^{\rm in}}^{z_{\rm S}}\,dz_{\rm S}' \,
{dN_{N_I}\over dz_{\rm S}} \,\exp\left[-\int_{z_{\rm S}^{\rm in}}^{z_{\rm S}'}\,dz_{\rm S}'' \, 
W_{\a}(z_{\rm S}'') \right]  \,  .
\ee
The wash-out is active (i.e., $W_{\a} > 1$) in an interval $z^\a_{\rm on} \gtrsim z_{\rm S} \gtrsim z^\a_{\rm off}$,
where $z^{\a}_{\rm on} \simeq 2/\sqrt{K_{1\a}+K_{2\a}} \ll 1$. For $z_{\rm S}\lesssim z_{\rm on}$ the wash-out can
be neglected and moreover decays are not fast enough to track the equilibrium distribution so that in this
regime $N_{N_I}-N_{N_I}^{\rm eq}\simeq z_{\rm S}^2/4$. Moreover, for $z_{\rm S} \ll 1$, one has
$D_I+S_I \simeq 0.2\,K_I$. In this case one obtains for the efficiency factor
\be\label{lowz}
k_{I\a}(z_{\rm S} \lesssim z^\a_{\rm on}, K_I) \simeq {0.2\,K_I \over 12}\,z_{\rm S}^3 \,  .
\ee 
On the other hand, for $z\gtrsim z_{\rm on}^{\a}$,  since we are in the strong wash-out regime 
with $K_{1}, K_{2} \gg 1$, one can
approximate $dN_{N_I}/dz \simeq dN^{\rm eq}_{N_I}/dz$ and, noticing that
\be
{dN^{\rm eq}_{N_{\rm S}}\over dz_{\rm S}} = -{2\,W_{\a}^{ID}\over (K_{1\a}+K_{2\a})\,z_{\rm S}} \,  ,
\ee
one finds a very simple analytical expression
\be\label{highz}
\kappa_{I\a}(z_{\rm S},K_{1\a}+K_{2\a}) \simeq {2 \over (K_{1\a}+K_{2\a})\,\bar{z} \, j(\bar{z})} \, 
\left(1 - e^{-{(K_{1\a}+K_{2\a})\,\bar{z} \, j(\bar{z}) \over 2}\,\left(1-N^{\rm eq}_{N_I}(z_{\rm S})\right)} \right) \,  ,
\ee
where $\bar{z}={\rm min}(z,z_B)$. At large values of $z_{\rm S} \gg z_{\rm in}$, this simply reduces to
$\kappa_{I\a}\simeq {2 / [(K_{1\a}+K_{2\a})\,\bar{z} \, j(\bar{z})]}$. 

A good interpolation between (\ref{lowz}) and $(\ref{highz})$, 
working at all values of $z_{\rm S}$, is then given by
\be
\kappa_{I\a}(z_{\rm S},K_I,K_{1\a}+K_{2\a}) \simeq 
{0.2\,K_I \over 12}\,z_{\rm S}^3 \, \left(1 + {0.2\,K_I\,(K_{1\a}+K_{2\a}) \over 24}\,z_{\rm S}^3\,\bar{z} \, j(\bar{z})\right)^{-1} \,  .
\ee
For the final value, using Eq.~(\ref{highz}), one has
\be
\kappa^{\rm f}_{\a}(K_{1\a}+K_{2\a}) \simeq {2 \over (K_{1\a}+K_{2\a})\,z_B \, j(z_B)} \, 
\left(1 - e^{-{(K_{1\a}+K_{2\a})\,z_B \, j(z_B) \over 2}}\right) \,   ,
\ee
where 
\be
z_B(K_{1\a}+K_{2\a}) \simeq 2 + 4\,(K_{1\a} + K_{2\a})^{0.13} \, e^{-2.5/(K_{1\a}+K_{2\a})} \,   .
\ee
Note that the final value is independent of $K_I$ and, therefore, it is the same for both RH neutrinos.
Since in the strong wash-out regime $z_B \gg 1$, one can approximate $j(z_B) \simeq 1$, meaning that
$\D L =1$ scatterings give a small correction that can be neglected with good approximation. This makes the solution
quite stable under the inclusion of many different subtle and complicated effects, like for example a precise calculation
of the thermal Higgs mass. Also let us notice that since we are in the fully three flavoured regime with masses 
$M_{\rm I} \ll 10^{9}\,{\rm GeV}$ and since, on the other hand, $M_I \gg 100\,{\rm GeV}$ and 
the asymmetry is produced from decays when RH neutrino mixing gives a negligible contribution,
then Boltzmann equations are expected to work very well and a density matrix formalism with
more heavy neutrino flavours \cite{lepdm} is not expected to be necessary to describe 
leptogenesis, though a dedicated analysis might be interesting to confirm this expectation.
%
%

\end{document}